\documentclass[useAMS,usenatbib]{mn2e}
\usepackage{mathtools} 
\usepackage{graphicx}
\usepackage{float}
\usepackage{lscape}

\newcommand{\tmax}{\ensuremath{t_{\mathrm{max}}}}
\newcommand{\trise}{\ensuremath{t_{\mathrm{rise}}}}

\newcommand{\tex}{\ensuremath{t_{\mathrm{ex}}}}
\newcommand{\rptf}{\ensuremath{R_{\mathrm{P48}}}}
\newcommand{\gptf}{\ensuremath{g_{\mathrm{P48}}}}
\newcommand{\grlsq}{\ensuremath{gr_{\mathrm{LSQ}}}}

\newcommand{\nickel}{\ensuremath{^{\mathrm{56}}}Ni\,}

\defcitealias{RTLOSS2011}{Ganeshalingham et al. (2011)}

\title[The Rising Light Curves of SNe Ia]{The Rising Light Curves of Type Ia Supernovae}
\author[R. E. Firth et al.]{R. E. Firth$^{1}$, M. Sullivan$^{1}$,  A. Gal-Yam$^{2}$, D. A. Howell$^{3}$, K. Maguire$^{4}$, P. Nugent$^{5,6}$,\newauthor A. L. Piro$^{7}$, C. Baltay$^8$, U. Feindt$^{9}$, E. Hadjiyksta$^8$, R. McKinnon$^8$, E. Ofek$^2$, \newauthor D. Rabinowitz$^8$, E. S. Walker$^8$\\
$^{1}$School of Physics and Astronomy, University of Southampton, Southampton, SO17 1BJ, UK; r.firth@soton.ac.uk\\
$^{2}$Benoziyo Center for Astrophysics, Weizmann Institute of Science, 76100 Rehovot, Israel\\
$^{3}$Las Cumbres Observatory Global Telescope Network, Goleta, CA 93117, USA\\
$^{4}$European Southern Observatory, Karl-Schwarzschild-Str. 2, 85748 Garching bei M\"unchen, Germany\\
$^{5}$Computational Cosmology Center, Lawrence Berkeley National Laboratory, 1 Cyclotron Road, Berkeley, CA 94720, USA\\
$^{6}$Department of Astronomy, University of California, Berkeley, CA 94720-3411, USA\\
$^{7}$Theoretical Astrophysics, California Institute of Technology, 1200 E California Boulevard, M/C 350-17, Pasadena, CA 91125, USA\\
$^{8}$Department of Physics, Yale University, New Haven, CT 06250-8121, USA\\
$^{9}$Physikalisches Institut, Universit\"at Bonn, Nu\ss allee 12, 53115 Bonn, Germany}

\begin{document}

\date{}

\maketitle

\label{firstpage}

\begin{abstract}
  We present an analysis of the early, rising light curves of 18 Type
  Ia supernovae (SNe Ia) discovered by the Palomar Transient Factory
  (PTF) and the La Silla-QUEST variability survey (LSQ). We fit these
  early data flux using a simple power-law ($f(t)=\alpha\times t^n$) to determine
  the time of first light ($t_{0}$), and hence the rise-time (\trise) from first light
  to peak luminosity, and the exponent of the power-law rise ($n$). We
  find a mean uncorrected rise time of $18.98\pm0.54$\,days, with
  individual SN rise-times ranging from $15.98$ to $24.7$ days. The exponent
  $n$ shows significant departures from the simple `fireball model' of
  $n=2$ (or $f(t) \propto t^2$) usually assumed in the literature. With a mean value of
  $n=2.44\pm0.13$, our data also show significant diversity from event to event. 
  This deviation has implications for the
  distribution of \nickel throughout the SN ejecta, with a higher
  index suggesting a lesser degree of \nickel mixing. The range of
  $n$ found also confirms that the \nickel distribution is not
  standard throughout the population of SNe Ia, in agreement with
  earlier work measuring such abundances through spectral modelling.
  We also show that the duration of the very early light curve,
  before the luminosity has reached half of its maximal value, does
  not correlate with the light curve shape or stretch used to 
  standardise SNe Ia in cosmological applications. This has
  implications for the cosmological fitting of SN Ia light curves.
\end{abstract}

\section{Introduction} 

Type Ia supernovae (SNe Ia) are bright stellar explosions that can be
standardised and used as distance indicators over cosmic scales.
Relative distances calculated using SNe Ia were used to uncover the
accelerating expansion of the universe \citep{OEFS1998,MOLH1999}, and,
more recently in the era of large surveys, sufficient accuracy has
been attained to enable precise cosmological measurements
\citep{2009ApJS..185...32K,SNLS32011,2012ApJ...746...85S,2014ApJ...795...44R,2014A&A...568A..22B}.

SNe Ia are thought to be the result of a thermonuclear explosion
of a carbon-oxygen (CO) white dwarf star as a result of mass transfer
to the white dwarf in a binary system. This is supported by recent observations placing
constrains on the radius of the progenitor, consistent with a WD 
\citep{11FE2011,2012ApJ...744L..17B}. Two basic scenarios for the progenitor 
systems are generally considered: single-degenerate (SD)
systems \citep{1973SD}, comprising a white dwarf accompanied by a less
evolved secondary, and double-degenerate (DD) systems \citep{1984DD}
with two white dwarfs. 
Other variations on these two scenarios include the detonation of a helium 
shell on a CO white dwarf that leads to core detonation (double detonation)
\citep{1994ApJ...423..371W,1995ApJ...452...62L,2014ApJ...785...61S}, the triggering of dynamical
 burning during the DD tidal disruption itself \citep{2012ApJ...747L..10P}, 
 and collisions between two white dwarfs in a triple system 
 \citep{2012arXiv1211.4584K,2013ApJ...778L..37K}.
Whatever the mechanism, as the progenitor white dwarf's mass increases
and approaches the Chandrasekhar mass, $M_{ch}$, carbon burning is ignited and a
runaway thermonuclear explosion results (in the collisional and double detonation cases, 
the total mass may not need to reach $M_{ch}$ due to additional compression forces). 
However, many of the exact physical details behind this picture are poorly
understood (see the recent review of \cite{2014ARA&A..52..107M}).

Studying SNe Ia just after their explosion is important for
understanding the physics of the ejected material. Immediately following the explosion, a shock travels through the envelope,
causing it to become unbound from the star. In the case where the shock is 
radiation-dominated, the shock travels outwards until the optical depth falls
to a level at which the radiation driving the shock can escape as a UV/X-ray flash.
The shock breakout of the explosion itself is likely too dim and fast to be
detectable for extragalactic events \citep{2012ApJ...757...35R,2012ApJ...747...88N},
but emission from the cooling ejecta heated by the shock could be detected.
This shock-heated cooling is predicted to be faint but should be best 
observed in UV and blue optical bands \citep{2013ApJ...769...67P}. Adding to the difficulty in 
detection, the timescale for this faint emission is very short given the small size of the progenitor star 
\citep{2010ApJ...708..598P,11FE2011,2012ApJ...744L..17B}.
 Other sources of very early
emission could trace the presence of companion stars. For instance, if
the ejecta collides with a companion it may cause a disruption and
re-heating of the ejecta, as well as blowing a hole in the ejecta,
where high energy emission could escape \citep{SCSC2010}. However, optical searches 
for this effect have so far been unsuccessful \citep{CSNP2011,SDDP2010}. Another
possible source of emission may arise from the SN Ia ejecta
interacting with shells of circumstellar material (CSM) previously
ejected by the system \citep{2007DCSM}. 

Interaction can occasionally be seen in the spectra
of a SN either through absorption or emission.
The strongest emission lines have been seen in SNe Ia initially mis-classified as 
type IIn SNe due to strong H$\alpha$ emission \citep{SilCSM2013} and have been
termed SNe Ia-CSM. 
This emission arises by conversion of of the kinetic energy of the fast-moving SN 
ejecta into radiation by shock interaction with a slow-moving CSM.

However, the bulk of the early optical light curve of a SN Ia is
powered by the radioactive decay of \nickel ($\rightarrow$ $^{56}$Co
$\rightarrow$ $^{56}$Fe) synthesized in the explosion
\citep{1960ApJ...132..565H,1969ApJ...157..623C,1982ApJ...253..785A,2000ApJ...530..744P},
and thus the shape of the light curve gives clues as to the
distribution of the \nickel in the ejecta. The first photons that
diffuse out of the ejecta result from energy deposition from the decay
of the \nickel that is located furthest out in the ejecta
\citep{2013ApJ...769...67P}. This process is not instantaneous, and as such, there 
may be a dark phase between explosion and first light, as has recently been implied by
abundance tomography \citep{2013MNRAS.429.2228H,2014MNRAS.439.1959M}.
The escape of the first photons starts the rise of the light
curve, and as the ejecta expands, photons generated by energy
deposited by deeper \nickel escape. The expanding ejecta become
less opaque, increasing the amount of energy escaping, and the point
at which the energy radiated is equal to the energy deposited by
\nickel is identifiable as a point of inflection on the light curve
\citep{2000ApJ...530..744P}. The ejecta continue to radiate
previously deposited energy as well as the energy instantaneously
deposited by ongoing \nickel decays, and consequently the peak of
the SN light curve occurs several days later.  The time between the
first photons escaping the ejecta (not necessarily the time of
explosion) and this peak is the `rise-time', with a value of $\sim17.5$
days to $B$-band peak for a normal SN Ia event \citep{SDDP2010}.

Although SNe Ia show considerable variation in their peak brightness
from event to event, they are ultimately standardisable (Phillips
1993) in the sense that brighter SNe Ia have slower evolving light
curves. This is usually parametrised by either a stretch-like
parameter \citep[e.g.][]{MOCP1997,2007A&A...466...11G}, often denoted $x_1$,
 which measures the speed of a SN Ia relative to a normal event, or a $\Delta m$-like
parameter, which measures the rate at which a light curve fades after
peak brightness \citep{AMTS1993,1996ApJ...473...88R}.

The width of the bolometric light-curve of a SN Ia is related to 
the photon diffusion time \citep{2000ApJ...530..757P,2007ApJ...662..487W}.
A photon emitted in a \nickel decay will random walk out of the ejecta, 
depositing energy at each collision. A longer diffusion time means that 
the photon spends longer within the ejecta and as such deposits more 
energy in total, both increasing the peak brightness and stretching the 
light curve. The important parameters for determining the bolometric
 diffusion time are the mass of the ejecta, the kinetic energy, the radial
 distribution of \nickel and the effective opacity \citep{2007ApJ...662..487W}. 
 The opacity increases with the ionisation state of Fe-group elements, which
 blanket the blue, and, as this increases with temperature, links opacity to the 
 \nickel mass - as hotter, brighter SNe Ia have more \nickel. 
 
 The study of SN Ia rise times has a long history
 \citep{1984SvA....28..658P,RTNS1999,RTSNLS2006,2007ApJ...671.1084S,RTSDSS2010,RTLOSS2011,RTSNLS2012}.
 \citet{1984SvA....28..658P} used 54 literature SNe Ia to demonstrate
 a range in rise time values, with the rise time correlating with the
 decline rate over 100 days. \citet{RTNS1999} used 30 unfiltered CCD
 observations and data at an earlier epoch than previously available,
 and measured $\trise=19.5\pm0.2$ days. They found that the rise time
 was correlated with peak luminosity in the sense that longer
 rise-times were found in brighter SNe, as expected if the speed of
 the early light curve correlates with the light curve shape.

 More recent work has used high-redshift SN surveys such as the
 Supernova Legacy Survey \citep[SNLS,][]{2006A&A...447...31A} and the
 Sloan Digital Sky Survey (SDSS) SN search
 \citep[][]{2008AJ....135..338F}. These surveys achieve a large sample
 size with very early detections due to their large search volume,
 time-dilated SN light curves, and high-cadence repeat imaging of
 `blank' areas of sky. Lower redshift surveys, such as the Lick
 Observatory Supernova Search \citep[LOSS,][]{2000AIPC..522..103L}
 that targeted nearby luminous galaxies, obtained a higher signal to
 noise, but located fewer SNe due to a smaller search volume. These
 surveys extract a rise time from their SNe by correcting all the SNe
 to the same peak brightness and light-curve width, and using a single
 rise-time value to represent the resulting distributions
 \citep{RTSNLS2006,RTSDSS2010,RTLOSS2011,RTSNLS2012}. These studies
 were also able to investigate the shape of the early light curve,
 parameterising the early luminosity evolution as a power-law with
 exponent $n$. They generally found values of $n$ consistent with 2
 (i.e., a light curve evolution proportional to $t^2$), and the results of
 these studies are summarised in Table~\ref{RT_table}.

 Some subtleties have emerged. Using eight well-sampled SNe Ia
 corrected for light curve width, \citet{2007ApJ...671.1084S} found a
 range of rise times with a dispersion of $0.96^{+0.52}_{-0.25}$ days,
 and some evidence for a bimodal distribution.  \cite{RTLOSS2011} and
 \cite{RTSDSS2010} use `2-stretch' models to fit stretches to the
 rising and falling sections of the light curves separately.
 \cite{RTLOSS2011} note that the rise time of high-stretch SNe Ia is
 shorter than would be expected based on the rest of their light curve
 shape.

 A handful of well-sampled, high-S/N local SNe Ia, in some cases
 discovered just a few hours after first light, have sufficient data
 to individually constrain the rise-time exponent ($n$): SN\,2011fe,
 $n=2.01\pm0.01$ \citep{11FE2011}; SN\,2010jn, $n=2.3\pm0.6$
 \citep{2013MNRAS.429.2228H}; SN\,2013dy, $n=2.24\pm0.08$
 \citep{2013ApJ...778L..15Z}; and SN\,2014J, $2.94\pm0.20$
 \citep{2014ApJ...783L..24Z,2014ApJ...784L..12G}. All of these studies
 found $n\geq2$. Such early-time data are particularly valuable for
 placing constraints on the progenitor \citep{11FE2011} and the
 physical processes within the ejecta \citep{2013ApJ...769...67P}.

\begin{table}{}
\centering
\caption{Rise-time results from the literature}
 \begin{tabular}{@{}lcccccc} 
 
  \hline
  Survey & n & Rise Time (days) &\\
  \hline
   SDSS$^1$ & $1.8^{+0.23}_{-0.18}$&$17.38\pm0.17$ & \\
   LOSS$^2$ & $2.2^{+0.27}_{-0.19}$ & $18.03\pm0.24$ & \\
   SNLS$^3$ & $1.92^{+0.31}_{-0.37}$ & $16.85^{+0.54}_{-0.81}$ &\\
  \hline
\label{RT_table}
 \end{tabular}
 \\ \small  1. \citet{RTSDSS2010}, 2. \citet{RTLOSS2011}, 3. \citet{RTSNLS2012}
\end{table}

In this paper, we use 18 SN Ia discoveries from two low redshift SN
surveys: the Palomar Transient Factory \citep{PTFS2009,EOTS2009}, and
the La Silla-QUEST Variability Survey \citep{2013PASP..125..683B}.
Both surveys operate with a similar 1-3\,day cadence, and are
wide-area rolling searches. This ensures that the early SN light
curves are well-sampled, with strong constraints on the SN first light.
 This also means that, rather than calculating an average
rise-time for the survey ensemble, individual events can be fitted
both for the rise-time and the exponent of the power-law rise. We
consider how the light curve behaves at these early times, what this
can tell us about the physical conditions in the ejecta, and how this
may relate to the progenitor.  We also investigate the subclass of SNe
Ia-CSM \cite{SilCSM2013} to establish to what extent their rises are
consistent with `normal' SNe Ia.

A plan of the paper follows. In Section~\ref{sec:data} we present the
SN Ia data used in our analysis and our sample of 18 SNe Ia.
Section~\ref{sec:methods} contains a review of the parameterisations
of the early time light curves of SNe Ia, and the methods applied to
fit them to the data.  Section~\ref{sec:results} presents the results
of our study of a sample of 18 `normal' SNe Ia, and in
Section~\ref{sec:discussion} these results are discussed, along with
SNe Ia-CSM.

\section{Data}
\label{sec:data}

\begin{figure}
	\includegraphics[width=84mm]{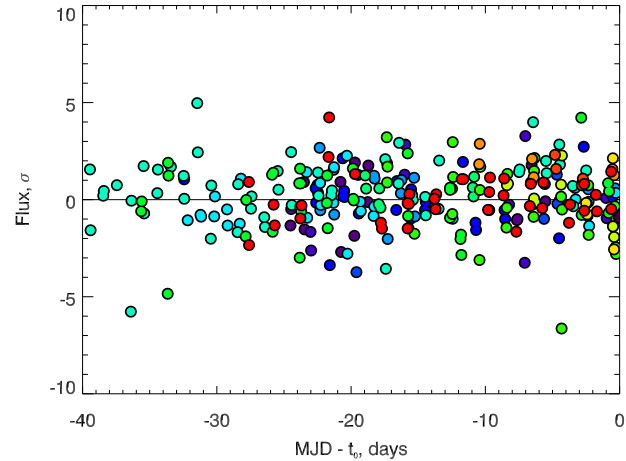}
	\caption{Flux prior to first light for our entire sample. No evidence of a systematic offset 
	is found. Each colour represents a different SN in our sample (online version only).}
	\label{pre_t0}
\end{figure}

\begin{figure*}
	\begin{center}
	\includegraphics[width=168mm]{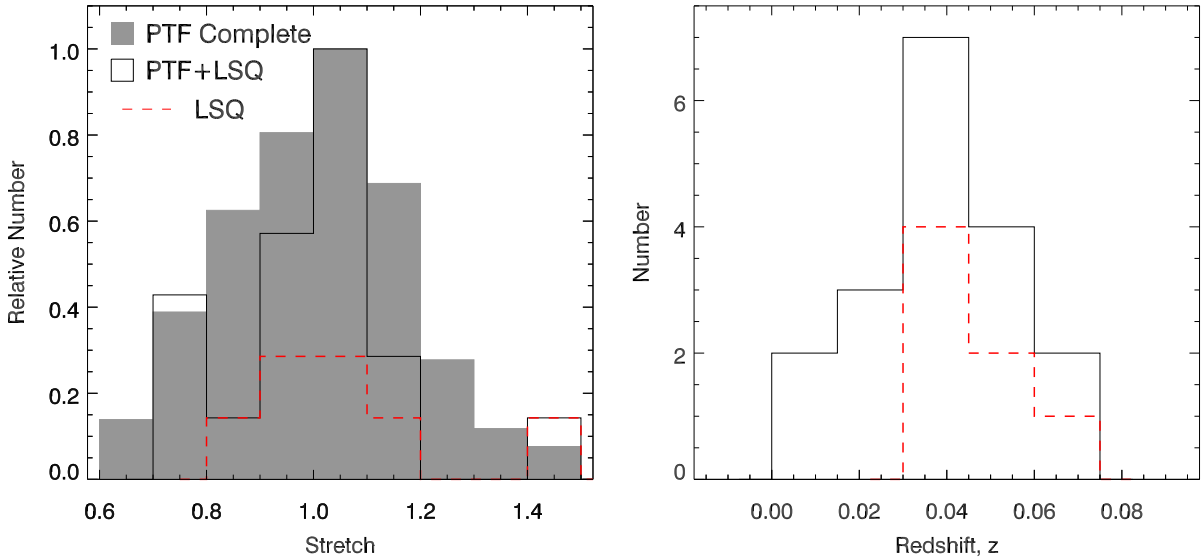}
	\end{center}
	\caption{\textit{Left Panel}:\,The stretch 
          distribution of our combined PTF and LSQ (black) and LSQ (red) data sets.
          The stretch was measured using SiFTO on data later than
          $\tau< -10$, as detailed in \ref{sec:sn-ia-rise}. The grey histogram shows the distribution of the full
          $z<0.09$ PTF sample from \citet{2014MNRAS.438.1391P}.
          \textit{Right Panel}:\, The redshift distribution of our combined
          sample.}
          \label{z+str}
\end{figure*}

In this section we introduce the sources of the SN data used in this
paper, the SN photometry, and the selection of the events that we use
for our analysis sample.

\subsection{The supernova surveys}
\label{sec:supernova-surveys}

Our data come from two local, rolling SN surveys. The first is the
the Palomar Transient Factory \citep[PTF;][]{PTFS2009,EOTS2009}, a
wide-field survey using the CFH12k camera mounted on the 48-inch
Samuel Oschin telescope at Palomar Observatory (the P48 telescope).
The survey operated primarily in an $R$-band filter (hereafter \rptf),
with occasional runs in a $g$ filter (\gptf) around new moon. The
cadence varied between a few hours and 5 days, although after selection cuts all the data
in this paper have a cadence of 4 days or better, and each 60s
exposure typically reached a depth of $\rptf\approx21$. The
combination of depth and cadence enabled the discovery of around 1250
spectroscopically confirmed SNe Ia (e.g., \citealt{2014MNRAS.438.1391P}).
The PTF images were processed by the PTF/IPAC pipeline described in 
\citet{2014PASP..126..674L} and are photometrically calibrated \citep{2012PASP..124...62O}

The second survey is the La Silla-QUEST variablity survey
\citep[LSQ;][]{2013PASP..125..683B}, a southern-hemisphere variability
survey using the 10-deg$^{2}$ QUEST instrument
\citep{2007PASP..119.1278B} on the 1.0m European Southern Observatoty
(ESO) Schmidt telescope at La Silla, Chile. LSQ operates with a
cadence of between 2 hrs and 2 days, using a broad $gr$ filter
(hereafter \grlsq).

The supernovae in this paper were spectroscopically confirmed using
 the Palomar Observatory Hale 200-in 
and the double spectrograph, the William Herschel Telescope (WHT) and 
the Intermediate dispersion Spectrograph and Image System (ISIS), 
the Keck-I telescope and the Low Resolution Imaging Spectrometer
 \citep[LRIS;][]{1995PASP..107..375O}, the Keck-II telescope and the 
DEep Imaging Multi-Object Spectrograph \citep[DEIMOS;][]{2003SPIE.4841.1657F},
the Lick Observatory 3m Shane telescope and the Kast Dual Channel Spectrograph
 \citep{miller1994kast}, the Gemini-N telescope and the Gemini
Multi-Object Spectrograph \citep[GMOS;][]{2004PASP..116..425H} and the University of Hawaii 88-in 
and the Supernova Integral Field Spectrograph \citep[SNIFS;][]{2004SPIE.5249..146L}.

All of the classification spectra for SNe in this paper are available 
via the WISeREP archive \citep{2012PASP..124..668Y}, and are also presented
in \cite{2014MNRAS.444.3258M}.
\subsection{SN photometry}
\label{sec:photometry}

Our photometric light curves originate from the
original SN searches. We use a single pipeline
written by one of us (MS) to construct all of the light curves from
both PTF and LSQ. This pipeline has been used extensively in earlier
PTF papers \citep[e.g.,][]{HSTS2012,2013Natur.494...65O,2014ApJ...781...42O,2014MNRAS.438.1391P} 
and we summarise the main details here.

The photometric pipeline runs on image subtraction, constructing a
deep reference image from data prior to the SN explosion, registering
this reference to each image containing the SN light, matching the
point spread functions (PSFs), performing image subtraction, and then
measuring the SN flux using PSF photometry on the difference images.
The PSF is determined using isolated stars in the unsubtracted images,
and the image subtraction uses a pixelized kernel (similar to
that in \citealt{2008MNRAS.386L..77B}). The SN position is measured from epochs when
the SN is present with the highest S/N (typically the position is
determined to better than 0.05-0.1 pixels), and then the PSF
photometry is performed in all images with this position fixed, avoiding
biases in the low S/N regime (see Appendix B of \cite{2007A&A...466...11G}\,for a
discussion).

The flux calibration is to the Sloan Digital Sky Survey \citep[SDSS;][]{2000AJ....120.1579Y} 
Data Release 10 \cite[DR10;][]{2014ApJS..211...17A} if the
SN lies within that survey's footprint, or otherwise to the photometric
catalogue of \citet{2012PASP..124..854O} for \rptf, or the AAVSO Photometric All-Sky Survey, APASS, 
\citep{2009AAS...21440702H}, for the other filters.

Our method is sensitive to variation in the data at very early times, 
i.e. very low flux levels, it is important to test for systematic effects. This was performed
by averaging the points before the SN first light. Prior to the explosion, 
the flux level is consistent with 0, with no evidence of a systematic offset, as shown in 
figure \ref{pre_t0}.

An example of the data used can be found in table \ref{data_tab}, and the entire dataset is 
available in online supplemental material.

\begin{figure}
	\includegraphics[width=84mm]{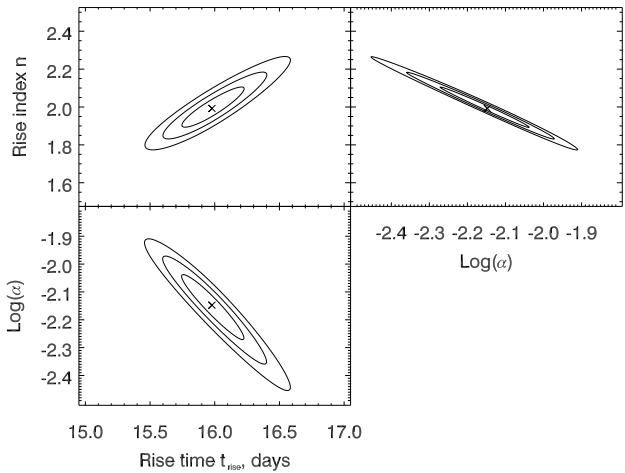}
	\caption{Probability distribution contours from the fit of SN
          PTF09dsy. Contours enclose 99.73\%, 95.45\%, and 68.37\% of
          the total probability.}
	\label{09dsycontours}
\end{figure}

\begin{figure*}
	\begin{center}
	\includegraphics[width=150mm]{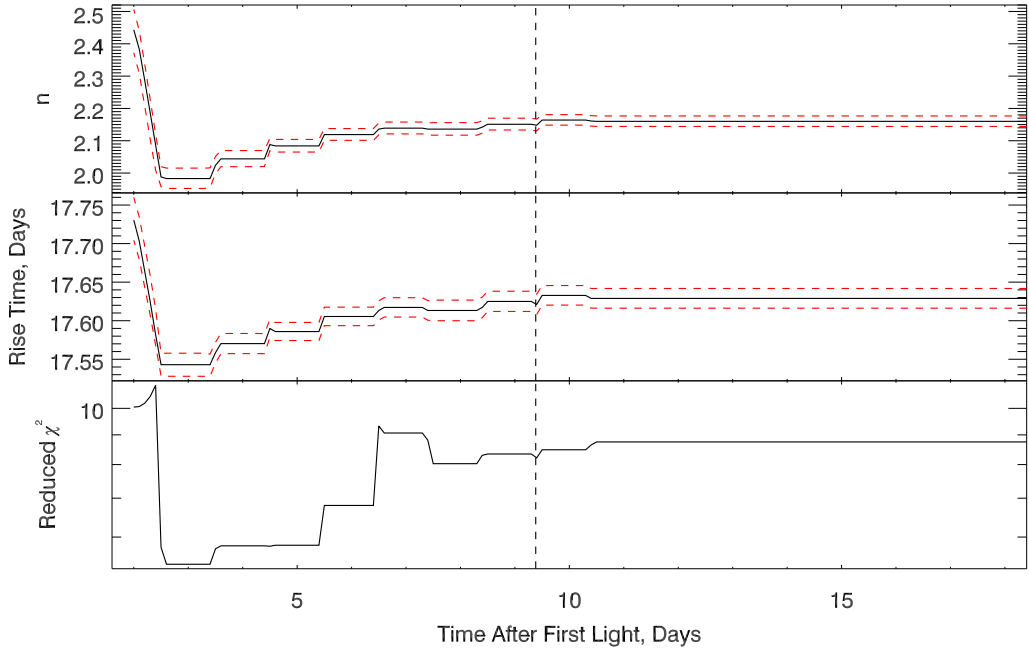}
	\end{center}
	\caption{The fit parameters for SN\,2011fe as a function of
          the epoch range over which the fit is performed, from
          2 to 18 days after first light. \textit{Top panel}: The
          change in $n$, \textit{Middle panel}: The variation in rise
          time, \textit{Bottom panel}: The evolution of the goodness
          of fit statistic, $\chi^{2}$ per degree of freedom, as
          defined in Equation~\ref{eq3}. The vertical dashed line
          shows the time at which the lightcurve reaches half of its
          maximum value ($t_{0.5}$)}
	\label{kly_cutoff}
\end{figure*}

\label{sec:light-curve-param}

We determine the light curve parameters for each SN Ia in our sample
using the SiFTO light curve fitter \citep{SiFTO2008}. SiFTO
manipulates the properties of a time-series SN Ia spectral energy
distribution (SED) in order to best fit an observed light curve,
returning the SN stretch ($s$), the time of maximum light in the
rest-frame $B$-band (\tmax), a peak magnitude, and a colour parameter
$c$ for SNe with data in more than one observed filter. We then define
all phases ($\tau$) in the SN light curve as relative to this
maximum-light, i.e.  $\tau=t-\tmax$, thus epochs prior to maximum
light have negative phases. We note that the use of SIFTO rather than
(e.g.) SALT2 \citep{2007A&A...466...11G} is not critical in this paper as we do
not make use of the peak magnitudes of the SNe, nor their Hubble
residuals. The SiFTO stretch and SALT2 equivalent ($x_1$) agree very
well for the same objects (e.g., \citealt{2010A&A...523A...7G}).

The spectral time-series template used by SiFTO assumes a $t^{2}$
photometric evolution in the $B$-band at phases $\tau\leq-10$
(equivalently 8-10 days post explosion for a normal SN Ia) due to a
lack of accurate early SN Ia photometric data at the time the SiFTO
package was written. Since in this paper we are primarily interested
in the behaviour of this early time data, we remove all data with
$\tau<-10$ when fitting with SiFTO. We use an iterative fitting
process to do this, first using all the data to estimate the $\tau = -10$
epoch, and then refitting with data earlier than this removed.

\subsection{Sample Selection}
\label{sec:sample-selection}

\begin{figure}
	\includegraphics[width=84mm]{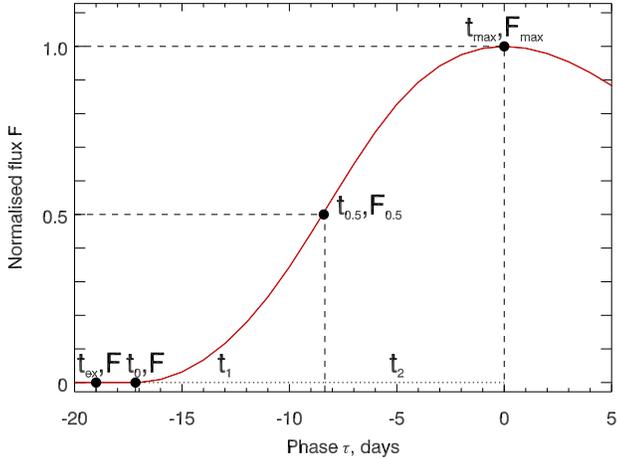}
	\caption{Schematic diagram of a SN Ia light-curve. The so called `dark
	phase' occurs between the time of explosion, $t_{ex}$ and the 
	time of first light, $t_0$. Also shown is the rise time split into its two 
	sections: $t_1$ is the region within which our fitting is performed, $t_2$
	is the time from the end of the fitting region up to maximum light}
	\label{labelled_template_color}
\end{figure}

As our study requires well-sampled and relatively high signal-to-noise
(S/N) data, there are several selection criteria that we make. We only allow SNe with both 
more than three epochs of data and more than 4 photometric points within 
the calculated fitting region (Section \ref{sec:defining-rise-time}, Fig. \ref{labelled_template_color}), 
as fewer would be insufficient to constrain the free parameters in the model.  
Light curves with more than four days between any two consecutive points 
are also excluded.

The distribution of stretch and redshift for our sample can be seen in
Fig.~\ref{z+str}. Using a Kolmogorov-Smirnov (KS) test, the stretch
distribution of our sample is consistent with being drawn from the
same distribution as the larger PTF sample with a probability of
$91\%$. Our mean redshift, $z=0.037$, is slightly lower than that of
the parent $z<0.09$ PTF sample, which has a mean redshift $z=0.056$.

\section{Analysis methods}
\label{sec:methods}

\begin{figure}
	\includegraphics[width=84mm]{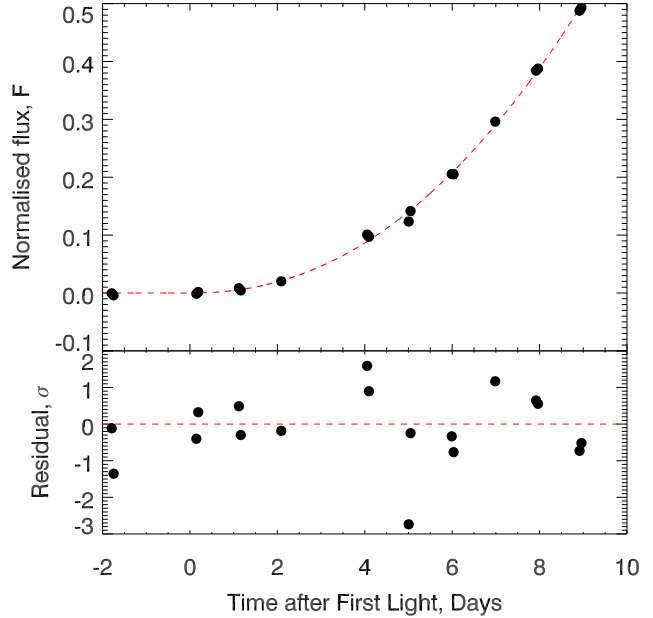}
	\caption{An example SN Ia lightcurve, best fitting model, and
          residual, for SN PTF11hub. The fit parameters can be found
          in Table~\ref{results_tab}. Uncertainties are plotted but are small, 
          residual is shown for clarity}
	\label{11hub_lc_res}
\end{figure}

We now turn to the analysis methods we will use in this paper, before
presenting the results in Section~\ref{sec:results}. We begin with a
discussion on the parameterisation used to fit the early portion of SN
Ia light curves.

\subsection{Rise Time Parameterisation}
\label{sec:rise-time-param}

The most widely used model parameterisation for the early-time SN Ia flux,
$f_{model}$, as a function of time, $t$, is

\begin{align}
f_{\textnormal{model}}(t)=\alpha(t-t_0)^n,
\label{eqn0}
\end{align}

\noindent for $t>t_0$, and 0 otherwise. Here, $\alpha$ is a normalising
coefficient, and $t_0$\ is usually treated as
the time of explosion. However, in this work we consider the possibility that
the time of the first photons escaping the ejecta, and the actual explosion of the 
SN, are distinct, and therefore we refer to $t_0$ as the time of first light.
Note that $t_0$ corresponds to the time at which the first photons leave the 
ejecta, which may differ from the time at which these photons can be 
detected by a given instrument. We do not require
that the join of this model at early times to the remainder of the SN
light curve be continuous.

In the cases of SN\,2011fe \citep{2014MNRAS.439.1959M} and SN\,2010jn 
\citep{2013MNRAS.429.2228H}, the precise measurement of the 
rise time led to some tension with 
spectral models, as the $t_0$ needed to match the observed 
abundances and spectral velocities is earlier than that
derived from the photometry. This implies that either the models are
incomplete in some way, or that there is a dark period between the
time of the explosion and the emergence of the first photons from the
ejecta. This is represented in Fig. \ref{labelled_template_color}, by the 
gap between $(t_{ex},F)$ and $(t_{0},F)$. 

Finally, $n$ is the index of the power-law.  The specific case of $n=2$ (giving a $t^2$
dependence) is known as the `expanding fireball' model, and is used
extensively in the literature as a reference model
\citep[e.g.,][]{RTNS1999,RTSNLS2006,2007ApJ...671.1084S,RTSDSS2010,RTLOSS2011,RTSNLS2012},
as it provides a good empirical match to early-time SN Ia
observations.

$\alpha$ is often ignored as a nuisance normalisation parameter, but
physically contains information about the mass, radius, \nickel, and
opacity of the ejecta.  \citet{2000ApJ...530..744P} show that the
(bolometric) rise time (\trise) depends on the same parameters as
$\alpha$, whilst $n$ is most sensitive to the mass and distribution of
\nickel and the shock velocity \citep{2013ApJ...769...67P}. As
$\alpha$ and $n$ depend on some of the same underlying physical
parameters and processes, degeneracies between them are expected.

The simple fireball model can be justified under the assumption of
constant photospheric temperature ($T$) and ejecta velocity ($v$)
\citep{RTNS1999}.  Assuming the emitting region is hot enough and the
SN is approximately represented by a black body, the standard optical
passbands lie in the Rayleigh-Jeans tail of the SN SED and the SN flux
will be $f\propto r^{2}T\propto v^{2}(t-\tex)^{2}T$, which for
constant $v$ and $T$ leaves $f\propto (t-\tex)^{2}$. It should be noted that if $T$ is constant, 
then the bolometric luminosity rises quadratically as well.  An assumption of
a $t^2$ rise was also shown by \cite{1982ApJ...253..785A} to be
reasonable, as the heating by radioactive decay should approximately
balance any adiabatic losses.

These underlying assumptions are, however, unrealistic over timescales
greater than a few hours. A more sophisticated treatment, following \citet{1982ApJ...253..785A}, is presented
in the analysis of SN\,2011fe in \citet{11FE2011}, who show that the $t^{2}$ relation is expected
without relying on the same assumptions. The rate of change of the internal energy can be
defined as a relationship between the energy deposited by \nickel,
the radiated luminosity and the internal radiation pressure.
Their method assumes that radiation pressure dominates and 
that the energy from \nickel is thermalised throughout the ejecta. Their
final, crucial assumption is that the elapsed time is much less than the \nickel\,
decay time, $t_{56}=8.8$ days. 
Despite the excellent fit to the data (\gptf, in this case of SN\,2011fe),
\citet{11FE2011} concede that this treatment is itself probably
simplistic in its analysis of the diffusion wave and distribution of
\nickel. If there is any colour evolution, then the bolometric light curve
will not be well fit by the same parameters.

This is developed further by \cite{RPRL2012} who predict
that the fireball model value of $n=2$ should be multiplied by a
coefficient related to the \nickel density gradient through the
ejecta and the shock velocity. This means that a single value 
of $n$ for all SNe Ia is not expected, and a range of $n$ is 
predicted instead.

Another method to probe both the structure of the ejecta and the 
assumptions is to investigate how the fit deviates from a $n=2$ fireball
model over time. The assumptions made in deriving the fireball model 
 \citep[outlined in][]{11FE2011} are strongest soon after explosion and 
 weaken at later times. As a result, in later sections, we consider a modified fireball model

\begin{align}
	f_{model}(t)= \alpha (t-t_0)^{n_0+ \dot{n}(t-t0)},
\label{ndot}
\end{align}

\noindent where $n_0$ is the rise index at time $t_0$ and  $\dot{n}$ 
is a variable measuring the deviation from the fireball model as time progresses. 

\subsubsection{Alternative Parameterisation}
A more recent parameterisation for the early light curve
uses a broken power law \citep{2013ApJ...778L..15Z}:
\begin{align}
	f_{model}(t)= \beta\left( \frac{t-t_{0}}{t_{b}}\right)^{\alpha1}\left[1+\left(\frac{t-t_{0}}{t_{b}}\right)^{s(n_1-n_2)}\right]^{-1/s},
\label{brokenpwr}
\end{align}

\noindent where $\beta$ is a normalisation constant, $t_0$ is the time of first
light, $t_{b}$ is the break time, $n_1$ and $n_2$ are the two rise
indices before and after the break, and $s$ is a smoothing parameter.
The motivation behind this approach is that changes in the index of a
power-law are a result of drastic changes in the temperature and
velocity of the fireball at very early times; the opposite of the
assumptions in the fireball model.
 An additional contribution may come
from the shock-heated cooling emission from the initial shock
breakout. To date, two SNe have been fitted with this model: SN\,2013dy 
\citep{2013ApJ...778L..15Z} and SN\,2014J 
\citep{2014ApJ...783L..24Z,2014ApJ...784L..12G}, in both cases 
predicting a faster rise time than that from a single power law model.

\subsection{Fitting Methods}
\label{sec:fitting-methods}

We perform fits of eqn (\ref{eqn0}) to our data, correcting for $1+z$ time dilation, using a grid-search
minimisation of the $\chi^{2}$ statistic over our three free parameters; 
$\alpha$, t$_0$ and $n$, i.e.,
\begin{align}
	\chi^{2}=\sum{\left( \frac{F-f_{model}}{\sigma_{F}}\right) ^2}
\label{eq3}
\end{align}

\noindent where $F$ and $\sigma_{\mathrm{F}}$ are the SN flux measurements and
uncertainties, $f_{\mathrm{model}}$ is the model SN flux from eqn (\ref{eqn0}),
 and the sum runs over all the data points. We compute probabilities over 
 a grid and report the mean value of the marginalised parameters as the
best-fits, with our quoted uncertainties enclosing 68.3\% of the
probability. The conversion from $\chi^2$ to probability, $P$, is
$P\propto e^{-\frac{\chi^{2}}{2}}$.

The grid size is chosen to enclose as close to 100\% of the
probability as is measurable. Specifically, 
we chose ranges of: $-15 < t_{0}-t_{0}(n=2)<10$, where , $0.0
< n < 8.0$ and $-9 <\log(\alpha)< 0$. An example of the range covered, and 
probability distribution, can be seen in
Fig.~\ref{09dsycontours}. We sample $\log(\alpha)$ rather than
$\alpha$ to better sample low values of $\alpha$, while maintaining
dynamic range.

This is a different, and slightly more direct approach to that used in
\citet{RTSNLS2006} and \citetalias{RTLOSS2011}, in which Monte Carlo
simulations are used to estimate the parameter uncertainties. However
those analyses were performed on stacked light-curve data (rather than
fitting individual objects), which requires a careful correction of
the light curve shape and SN flux normalisation. This can introduce
covariances between stacked data points, which demands a more
sophisticated Monte Carlo like approach to handle these covariances.

The ellipticity of the contours in Fig.~\ref{09dsycontours}
demonstrate the covariance between the parameters in a typical fit.
The strongest is found between $n$ and $\log(\alpha)$ but is present in
significant strength between all of the variables. 

\subsection{Defining the Rise-time region}
\label{sec:defining-rise-time}

The simple rise-time model of eqn.~(\ref{eqn0}) will only hold over
the first few days of the SN evolution, as at some epoch the rise of
the SN slows and eventually reaches a maximum point. Thus our first
task is to determine over which range the model holds, and thus
over which range we can fit data. 

Both \citet{RTSNLS2006} and \citet{RTLOSS2011} define the rise-time
region as earlier than 10 days before $B$-band maximum light, (i.e.,
$\tau<-10$). This may occur at a different number of days
post-explosion for different SNe Ia due to the stretching of the SN
light curves. As we are not stretch-correcting the raw data in this
study, prior to fitting, we instead prefer a definition relative to $t_0$.

We first fit the fireball model with an initial rise time region of $\tau<-10$.
This gives a first estimate of $t_0$ with $n=2$.
We then re-fit the light curve using data ranging from 2 to 18 days after $t_0$,
 and record the values of $n$ and $\trise$,
and the $\chi^2$ (Fig.~\ref{kly_cutoff}). The discontinuous jumps in
Fig.~\ref{kly_cutoff} are due to the inclusion of more data as the 
epoch range expands.  

The choice of the fitting region must balance two competing
constraints: there must be sufficient data to allow a meaningful
rise-time fit, yet the fitting region must not reach too far into the
photometric evolution where the rise-time parameterisation does not
hold. Balancing these requirements
across the sample, as well as taking into account the stability of the
result is challenging.

In many cases, the cut-off time that best satisfied these constraints was
nearly coincident with $t_{0.5}$, the time at which the SiFTO light
curve was at half of its maximum value (see Fig. \ref{labelled_template_color}).
$t_{0.5}$ is not reliant on either the stretch or a fixed number of
days, so is an ideal choice as a limit of the fitting region. It 
is also broadly consistent with $\tau<-10$ if the light curve
was stretch corrected. To ensure a consistent definition of $t_{0.5}$, after fitting, the value of
$t_{0.5}$ is re-calculated using the best fit to the data. This is in
most cases almost indistinguishable from that calculated from SiFTO,
indicating a good match, but is free of any reliance on the later time
data. Figure \ref{11hub_lc_res} shows the outcome of
this process, a best fit to PTF\,11hub.

This approach is similar to that used in \citet{RTSNLS2012}, where
they define their limiting epoch as the `transition phase',
$\tau_{t}$, where the light curve transitions from the rise to
the main body, and they find $-10\la\tau_{t}\la-8$.

Following this procedure to define our cut off, our final fit for each SN was
 performed on the data where $t_{0}(n=2)-2<t<t_{0.5}$ was satisfied; that
 is, we fit data up to two days before the time of first light, as calculated from an
 enforced fireball $n=2$ fit, and less than half of the maximum brightness.
  Imposing the lower limit was found not to affect the outcome of the fits,
 but removed contamination of the probability distribution from non detections.

\subsection{Fits to bolometric versus filtered data}
\label{sec:fits-bolom-vers}

Our next task is to establish the validity of comparing fits obtained
in different filters and at different redshifts, both for comparison to earlier work, and for
comparison between the different surveys in our sample.
Ideally we would measure the rise-time on the bolometric output of the
SN, but such data are not available, and hence we need to examine any
biases that might result; essentially, we are testing the effect of k-corrections. 
We test this by using the very well observed nearby SN Ia SN\,2011fe, which has
significant spectroscopic and photometric early time data.

We use 15 available pre-maximum spectra from the literature of
SN\,2011fe \citep{11FE2011,2012ApJ...752L..26P,2013A&A...554A..27P,2014MNRAS.439.1959M}.
 We measure synthetic light curves from
these spectra in the $B$, $V$, \gptf, \rptf, and \grlsq\ filters, as
well as a `pseudo-bolometric' band with a wavelength range
$3500-9000$\AA, with each spectrum scaled so that its synthetic \gptf\
magnitude matched that measured from the real \gptf\ photometry. Using
the spectral templates of \citet{2007ApJ...663.1187H}, this
pseudo-bolometric filter contains $\simeq70\%$ of the bolometric flux
at $t_{0.5}$ (corresponding to $\tau=-8.9$), and $\simeq72\%$ of the
flux at maximum light. The wavelength range of the pseudo-bolometric filter
was chosen as it is covered by most of our available spectra.

The uncertainties in our synthetic light curves come from the \gptf\
photometric uncertainties, with an additional systematic uncertainty
added in quadrature arising from relative flux calibration errors
(e..g, differential slit losses). We estimate this to be 1\%. These
synthetic light curves were then fit as described in
Section~\ref{sec:fitting-methods}, and the results are in
Table~\ref{results_11fe}.

\subsubsection{The Effect of Different Filters}
\label{sec:sn2011fe-spectr-test}

\begin{figure*}
	\includegraphics[width=150mm]{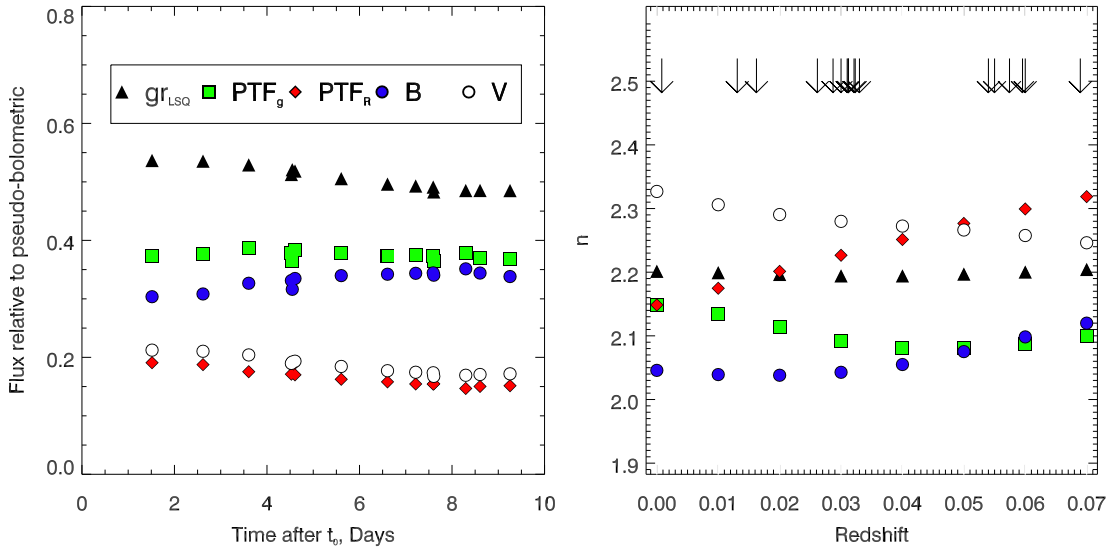}
	\caption{\textit{Left Panel}:The flux of SN\,2011fe through various filters
          relative to a pseudo-bolometric flux (see
          Section~\ref{sec:fits-bolom-vers} for details) as a function
          of epoch.  The filters are (top to bottom) \grlsq, \gptf,
          $B$, $V$, and \rptf\ filters.
          \textit{Right Panel}: The fits to the redshifted spectra of SN2011fe, over the range of 
          redshifts covered by our sample. The scatter around the mean redshift is consistent
          with the values found in the neighbouring panel. The arrows at the top of the figure show
           where our supernovae lie.}
	\label{bolfrac}
\end{figure*}

\begin{table}{}
\centering
 \caption{Results for the SN\,2011fe synthetic light curves. $\Delta n$ is the difference
 between n in a given filter and that of the pseudo-bolometric, i.e. $n^{filter}-n^{Pseudo-bol}$. Similarly, 
 $\Delta t$ is the difference between t nResults in \gptf\, differ
 from \citet{11FE2011} as a result of fitting a longer segment of the light curve.}
 \label{results_11fe}
 \begin{tabular}{@{}lccccc}
  \hline
 Filter & n & $\Delta$ n & \trise, & $\Delta\trise$\\
 & & & days & days& \\
  \hline
  Pseudo-bol & $2.23$ & $0.0$ & $17.75$ & $0.0$ \\
  $B$ & $2.05$ & $-0.18$ & $17.13$ & $-0.62$\\
  $V$ & $2.33$ &$0.1$& $18.54$ & $0.79$\\
  \gptf & $2.15$ &$-0.08$& $17.62$ & $-0.13$\\
  \rptf & $2.15$ &$-0.08$& $18.14$ & $0.39$\\
  \grlsq & $2.20$ &$-0.03$& $17.94$ & $0.19$\\
 \hline
 \end{tabular}
\end{table}

The results show some differences between different filters, due to
rapid evolution in the spectral features in each band.
Fig.~\ref{bolfrac} shows how the flux in each filter, relative to
pseudo-bolometric flux at that epoch, changes with time.  The \gptf\
band shows an almost constant flux ratio, but \rptf\ decreases with
time, while the $B$-band increases. 
The pseudo-bolometric value, $2.23$ is greater than that in \gptf, and greater than 2, 
in agreement with the findings of \citet{2014ApJ...784...85P}.
The broadest filter, \grlsq, is obviously the closest to bolometric,
but also shows a decreasing flux ratio. 
Note that earlier work has predominantly used data either in, or
corrected to, the $B$-band. Table~\ref{results_11fe} shows that this
fit has an $n$ closest to 2; however, it is significantly lower than
the values in the other filters, and is not consistent with the
pseudo-bolometric value.

In an attempt to further understand the colour evolution and its effect on $n$, templates
 from SALT2 \citep{2007A&A...466...11G} and \cite{2007ApJ...663.1187H}/SiFTO\citep{SiFTO2008}
were analysed, however, the nature of the investigation probes the very earliest epochs, 
where there have been few spectral observations. For example, 
the earliest \cite{2007ApJ...663.1187H} templates pre-maximum are based 
on 6 spectra with an average epoch of $\tau = -11.6$ days. 
Because of this, and despite being a single example, the data from SN\,2011fe are the 
best resource available at very early times. Comparing with SN\,2013dy \citep{2013ApJ...778L..15Z}, the colour 
evolution is similar, although SN\,2011fe exhibits a stronger Ca II IR triplet compared with
SN\,2013dy within the first 2 days. This does not fall within any of our filters, but falls within the 
pseudo-bolometric range.

In summary, by considering the different fits to our simulated photometry, 
we estimate the systematic effect of using different filters to be $n\pm0.1$

\subsubsection{The Effect of Redshift}
\label{sec:effect-z}

As our sample lies across a range of redshifts, we also need to ascertain the impact
that this has on the colour evolution. To do this, we performed the same procedure 
as in section \ref{sec:sn2011fe-spectr-test}, redshifting the spectra each time, 
up to our maximum redshift of $z=0.07$ before fitting. The results of these tests can be seen in the 
right-hand panel of figure \ref{bolfrac}.

We find that the broad \grlsq~filter provides data that is very stable with redshift, the value of $n$
remaining almost static. Leaving the rest-frame, the measured $n$ in \rptf~and \gptf~diverges,
with the \rptf~increasing and the \gptf~decreasing more steadily. The scatter around the mean redshift
 ($z=0.036$) is consistent with the dispersion between filters in table \ref{results_11fe}.
 
 In summary, we estimate the systematic effect of redshift to be reflective of that from using different filters,
 since the low redshifts do not shift the spectra by more than the width of our filters, that is, a systematic effect of at most
  $n\pm0.1$
 
\subsubsection{The Effect of Extinction}
 \label{sec:effect-ext}
 
We also performed these checks after reddening the spectra by $E(B-V)=1$ and found no 
 significant deviation in results.
 Additionally, inspecting the spectra at maximum light of each PTF object in
our sample, we find evidence of Na I D absorption
in only two SNe (PTF11gdh
 and PTF12gdq). These are discussed in Section \ref{sec:results} but
do not appear to be unusual events.

\begin{figure*}
	\includegraphics[width=150mm]{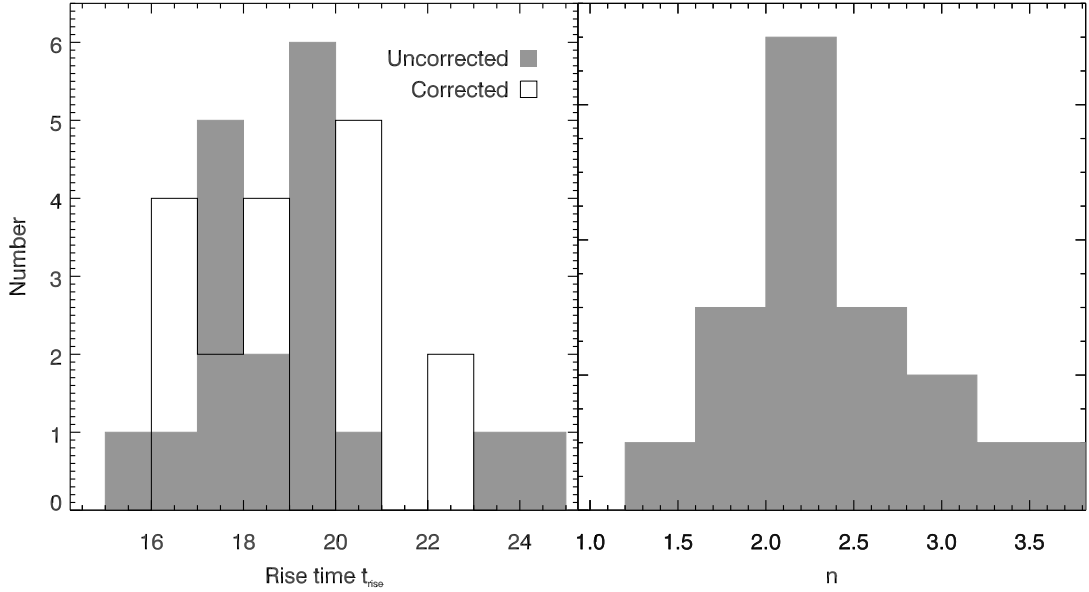}
	\caption{
          \textit{Left Panel}: Histogram showing the distribution of rise times for
          our sample leaving $n$ as a free parameter. The solid grey
          histogram shows the raw rise-time, while the unshaded
          histogram shows the rise-time corrected for stretch.
          \textit{Right Panel}: The histogram of the best-fit rise-index ($n$) values.
          The distribution has $n_{mean}=2.36\pm0.13$ and $n_{median} = 2.31$ 
          with a slight positive skew.
          }
	\label{tr+n_histo}
\end{figure*}

\section{Results}
\label{sec:results}

We now present the results of applying our fitting methods
(Section~\ref{sec:fitting-methods}) to our data sample
(Section~\ref{sec:sample-selection}). We first discuss the rise-time
analysis, followed by the rise index. Our results can be found in
Table~\ref{results_tab}.

\subsection{The SN Ia rise time}
\label{sec:sn-ia-rise}

The average rise-time of the 18 SNe Ia in our sample with $n$ a free
parameter in the fit, is $\trise = 18.98\pm0.54$\,days, or $\trise = 18.97\pm0.44$\,days
if the rise-times are stretch-corrected, where the uncertainties in both 
cases are the standard error on the mean (We exclude one SN, PTF12emp,
from this latter calculation as there is insufficient data to
reliably estimate a stretch.) For the stretch correction, we use
SIFTO to measure the stretch based on photometry later than
$\tau=-10$, and so it is independent of the shape of the early light
curve.  These values are longer than those found in previous work.
Assuming $n=2$, the mean rise-times are $\trise = 17.86\pm0.42$\,days
uncorrected and $\trise = 17.90\pm0.33$\,days, after stretch correction. The $n=2$ rise times
are shorter in both cases. These values are consistent with \cite{RTLOSS2011}, 
but lower than \cite{RTSNLS2006} and higher than those found in both 
\cite{RTSDSS2010} and \cite{RTSNLS2012} by $3\sigma$.

\begin{figure}
	\includegraphics[width=84mm]{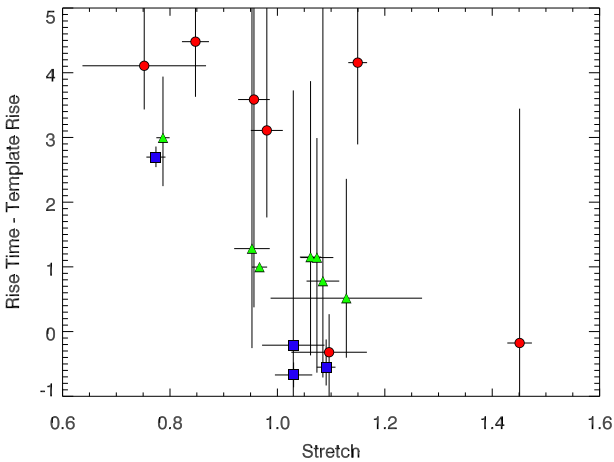}
	\caption{The difference between the measured \trise\ and the
          \trise\ expected from our SN template based on the fit
          stretch, plotted against the stretch. The coloured points
          denotes a binning by the rise index. Red circles are
          $n>2.4$, green triangles $1.9<n<2.4$, and blue squares
          $n<1.9$. Higher stretch SNe Ia have a \trise\ that is
         faster than that implied by the stretch-corrected template
          \trise.}
	\label{str_vs_delta_tr}
\end{figure}

\begin{figure}
	\includegraphics[width=84mm]{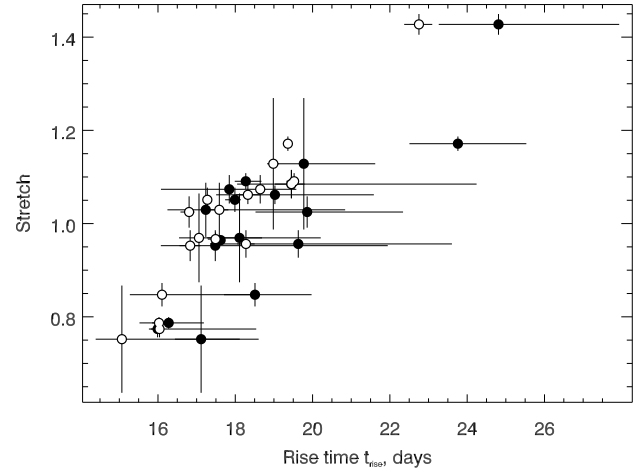}
	\caption{Stretch vs rise time. The black points are those
          fitted with a free $n$ parameter, the hollow points are
          those where the fitting has been constrained to $n=2$.          }
	\label{stretch_vs_tr}
\end{figure}

\begin{figure}
	\includegraphics[width=84mm]{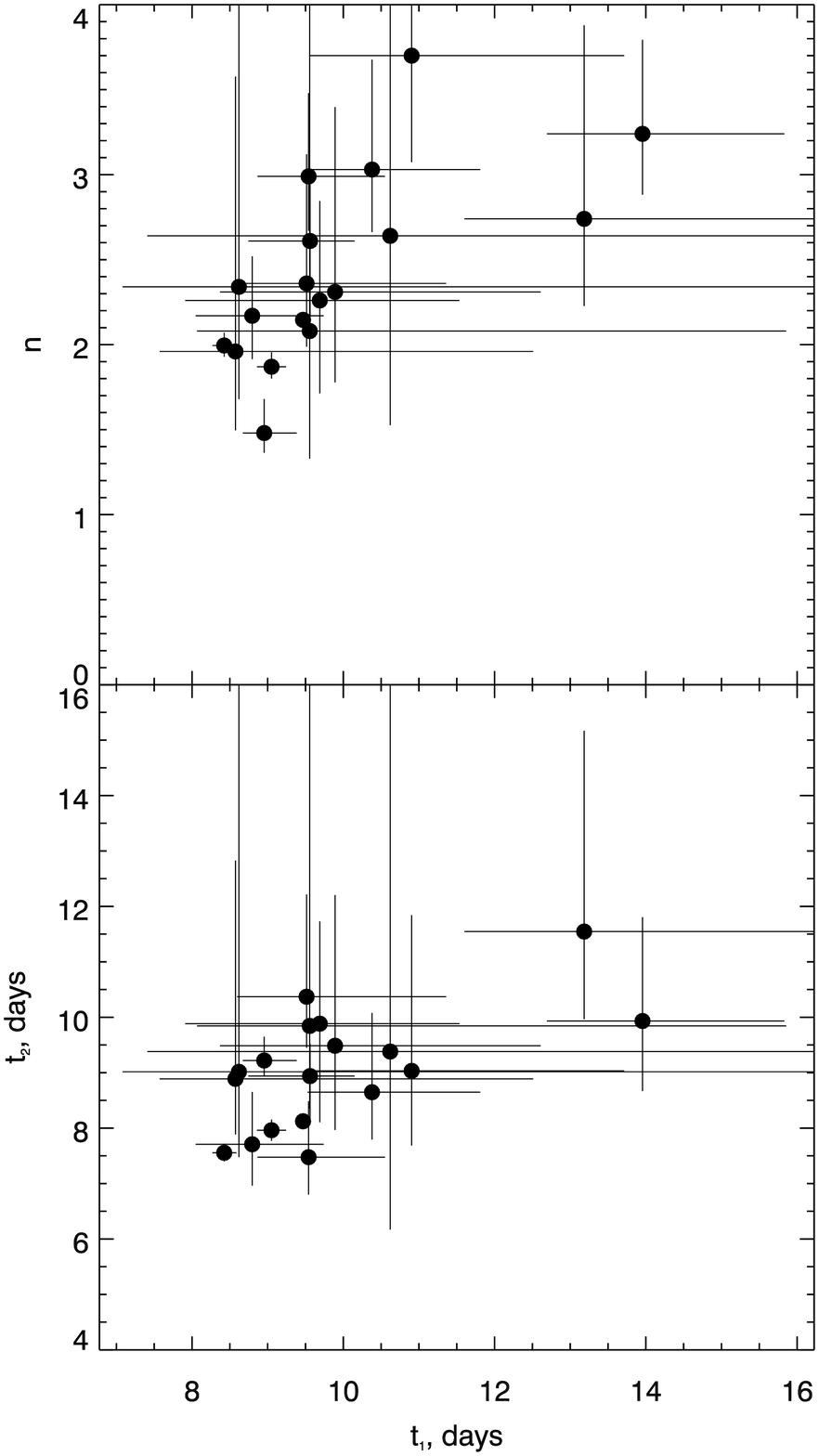}
	\caption{\textit{Top Panel}:\, $n$ vs $t_{1}$, in the region fitted, 
	$n$ correlates with the length of time from $t_0$ to $t_{0.5}$.
	 \textit{Bottom Panel}:\,$t_{2}$ vs $t_{1}$, no strong correlation can be seen 
	 between the early $t_1$ and late $t_2$ parts of the light curve.
	}
	\label{n_v_t2_t1}
\end{figure}

The histogram of the rise-time distributions can be found in
Fig.~\ref{tr+n_histo}. \citet{2007ApJ...671.1084S} previously
suggested that there may be two rise time modes, once the rise 
times have been corrected for the overall shape of the light curve 
(using the fall time). We do not find any evidence for this using 
either $n=2$ or $n$-free.

\citet{RTSDSS2010} and \citet{RTLOSS2011} both find that the fraction of
their sample that are slowest to decline after peak are amongst the
fastest to rise. Both studies therefore parameterise the width of the
light curve using two stretch parameters, one pre- and one
post-maximum.  \citet{RTLOSS2011} also find that the luminous SNe Ia
have a faster rise than expected based on a single stretch value. We
see a similar trend in Fig.~\ref{str_vs_delta_tr}, also lower stretch SNe 
appear to have slower light curves than would be expected from a single
stretch.

\begin{figure*}
	\includegraphics[width=150mm]{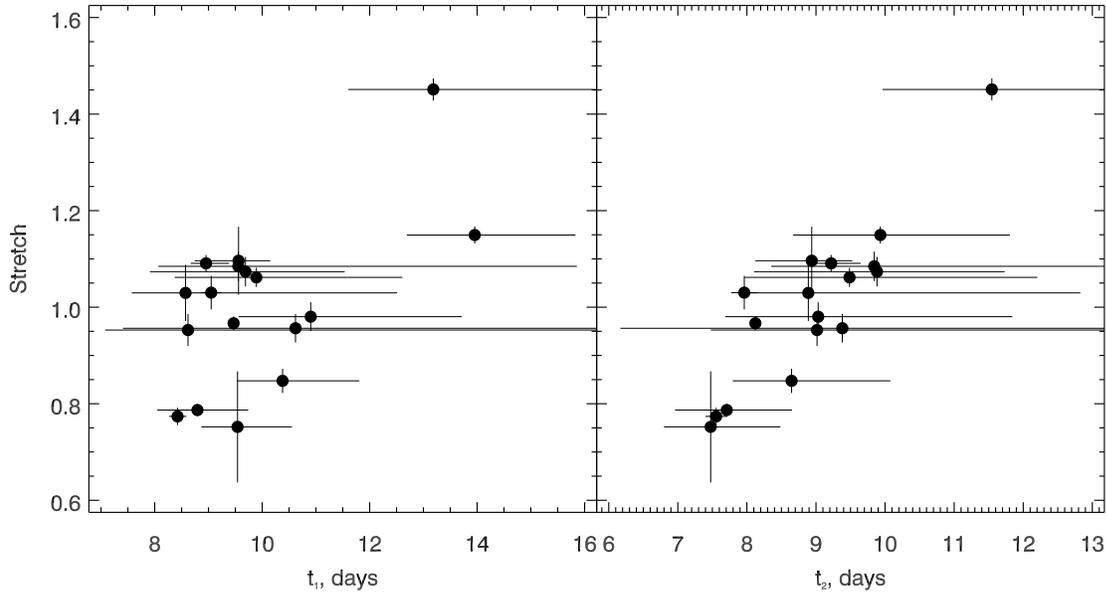}
	\caption{\textit{Left Panel}:\, Stretch vs $t_{1}$, no strong correlation can be 
	seen.
	\textit{Right Panel}:\, Stretch vs $t_{2}$, a clear correlation is visible.
	This is expected as the stretch was measured using data within this region, the strength of 
	the correlation is slightly weaker than expected.}
	\label{str_vs_t1_t2}
\end{figure*}

In Fig.~\ref{stretch_vs_tr}, we show the relation between stretch (again
calculated without the very early photometric data) and \trise. A
correlation is expected and observed in the data. The results are also 
shown for when $n = 2$ showing, on average, shorter rise times. 

The rise-time can be decoupled into two components: $t_{1}$, the time
between first light ($t_0$) and the time of half maximum
($t_{0.5}$), and $t_{2}$, the time between $t_{0.5}$ and \tmax ~
(shown in Fig. \ref{labelled_template_color}).
As can be seen in Fig. \ref{n_v_t2_t1}, surprisingly, these two
timescales do not show a particularly strong dependence, having a
 Pearsons Correlation Coefficient $P = 0.61$, and when imposing a stretch 
 cut commonly used in cosmology, $0.7\leq s \leq 1.3$ \citep{2011ApJS..192....1C}, 
which excludes LSQ12gpw, this drops to $0.43$. As the SiFTO fit includes data 
from $\tau>10$, which is roughly consistent with $t_{0.5}$, it is unsurprising 
that there is a strong correlation between $t_{2}$ and stretch (Fig. \ref{str_vs_t1_t2}),
 with $P = 0.89$. However, there is also no strong relationship between stretch
  and $t_{1}$ (Fig. \ref{str_vs_t1_t2}) ($P=0.57$, 
 which weakens to $P=0.34$ when imposing a stretch cut).

\begin{figure}
	\includegraphics[width=84mm]{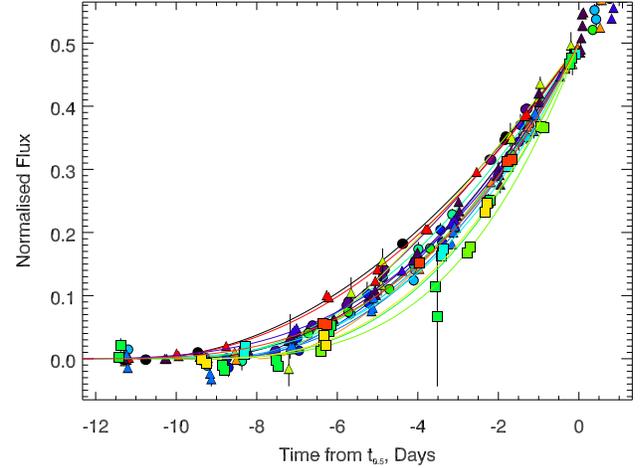}
	\caption{ Our sample of 17 SNe Ia (PTF12emp excluded, see Sec. \ref{sec:sn-ia-rise}), normalised, 
	stretch corrected and shifted to have coincident $t_{0.5}$. Note the diversity in the 
	early SNe Ia light curves even after stretch correction}
	\label{STACK_LC}
\end{figure}

The diversity of the early time light curves in our sample  can be seen in Fig.~\ref{STACK_LC}.
The light curves have been stretch corrected and normalised, and shifted to have a 
coincident $t_{0.5}$. Whilst when stretch corrected, in the $t_2$ distribution the scatter is reduced,
the data in $t_1$ still show a large amount of variation (Fig. \ref{STACK_LC}). 
This variation, even after stretch correction, may have been lost within instrumental noise in previous 
  surveys. To avoid introducing additional systematics due to misinterpreting this scatter,
  care must be taken when using SNe Ia data in this region for cosmology 
  as the variation is significant.

\subsection{The Rise Index - `$n$'}
\label{sec:rise-index-n}

\begin{figure}
	\includegraphics[width=84mm]{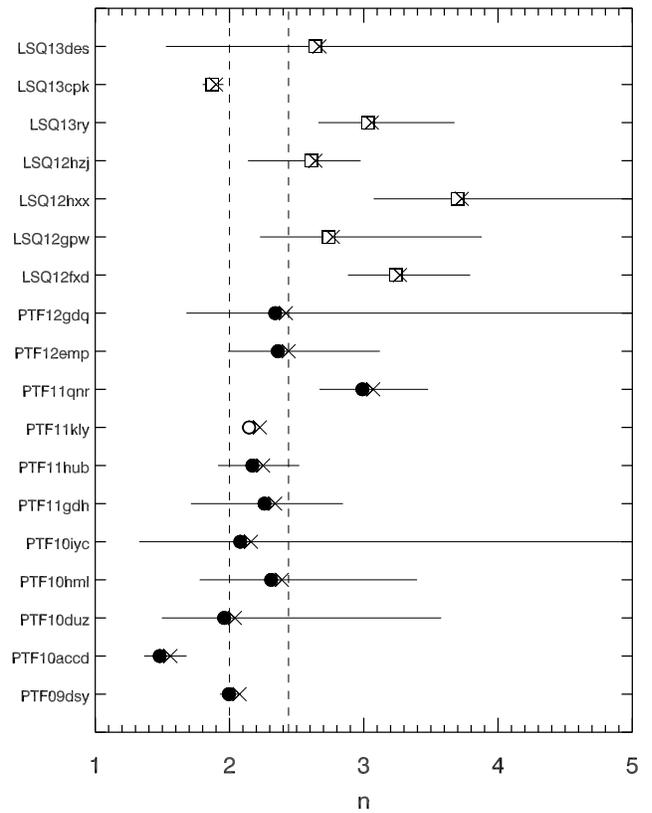}
	\caption{The best-fit `$n$' and uncertainty for
          each SN in the sample. Hollow squares are \grlsq observations, 
          solid circles are \rptf and hollow circles are \gptf. The dotted lines indicate $n=2$ and
          the mean of the sample, $n_{mean}=2.44\pm0.13$. The crosses 
          show the location of the points corrected to the `bolometric'
          value of $n$, if the SN\,2011fe correction (Table \ref{results_11fe}) holds 
          for other SNe }
	\label{n_res}
\end{figure}

\begin{figure}
	\includegraphics[width=84mm]{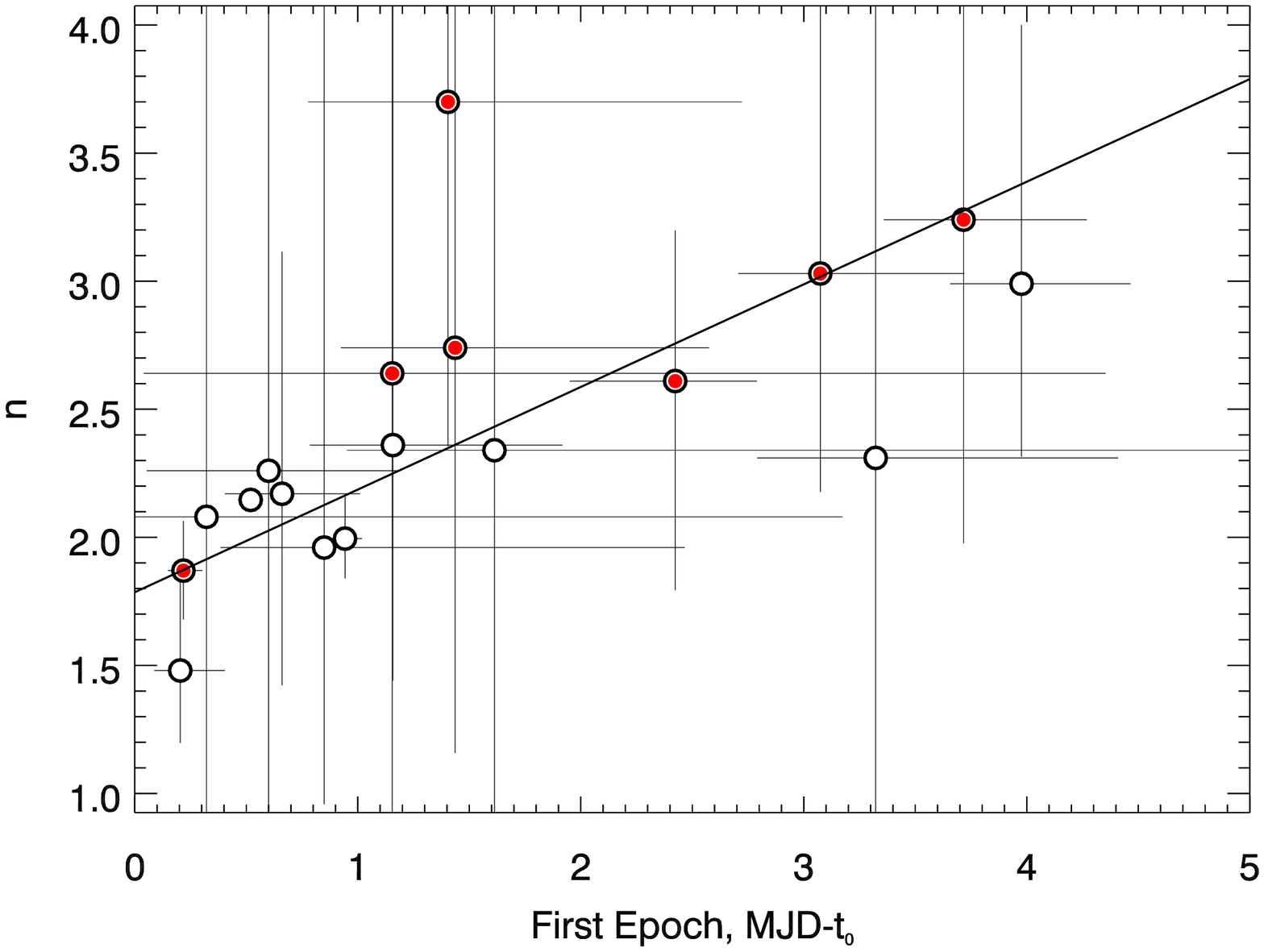}
	\caption{The fitted n plotted against the time of the first observation, relative to $t_0$ corrected
	for redshift.
	 LSQ SNe are shown in red (filled) and PTF are shown in white (hollow). The best fit to the data is overplotted.}
	\label{n_vs_obsdate}
\end{figure}

The distribution of the $n$ parameter, which can be seen in Fig.~\ref{tr+n_histo}, has
a mean of $n=2.44\pm0.13$ and a tail in the distribution towards
higher $n$. When corrected to a pseudo-bolometric value, as discussed in Section \ref{sec:fits-bolom-vers}, 
this becomes $n=2.50\pm0.13$. Both mean values, corrected and uncorrected, are not consistent with the $n=2$ fireball
model, although individual SNe Ia within the sample are consistent
with $n=2$ (Table~\ref{results_tab}); the $n$ values broken by SN name
are shown in Fig.~\ref{n_res}. To compare with previous work (Table \ref{RT_table}),
our $n$ value is marginally
consistent with \citet{RTLOSS2011}, who use the low-redshift LOSS sample and find
$n=2.2^{+0.27}_{-0.19}$. Our result is inconsistent with more recent
higher redshift studies, \citet{RTSDSS2010} or \citet{RTSNLS2012}.
Furthermore, the recent study of SN\,2014J has yielded a rise index of
$n=2.94\pm0.20$ \citep{2014ApJ...783L..24Z}. This lends further
evidence that there is not only a range of $n$, but that the centre of
the distribution is located at values $n>2$.
This result supports the finding of \citet{2014ApJ...784...85P} that a 
$t^{2}$ rise is not a generic property of SNe Ia.

As mentioned in Section \ref{sec:effect-ext}, we see non-zero Na I D absorption
 lines at the position of the host galaxy in the low-resolution spectra in in two SNe; 
 PTF11gdh and PTF12gdq. However, we cannot tell whether this is from the host 
 or from CSM interaction but the measurements of $n$ and \trise ~are not different from the bulk of the sample.
We do not find these SNe to occupy any unusual position in any of the parameter
space we investigate. Both have a value of n that is consistent with the mean value
within the calculated uncertainties.

\begin{figure*}
	\includegraphics[width=150mm]{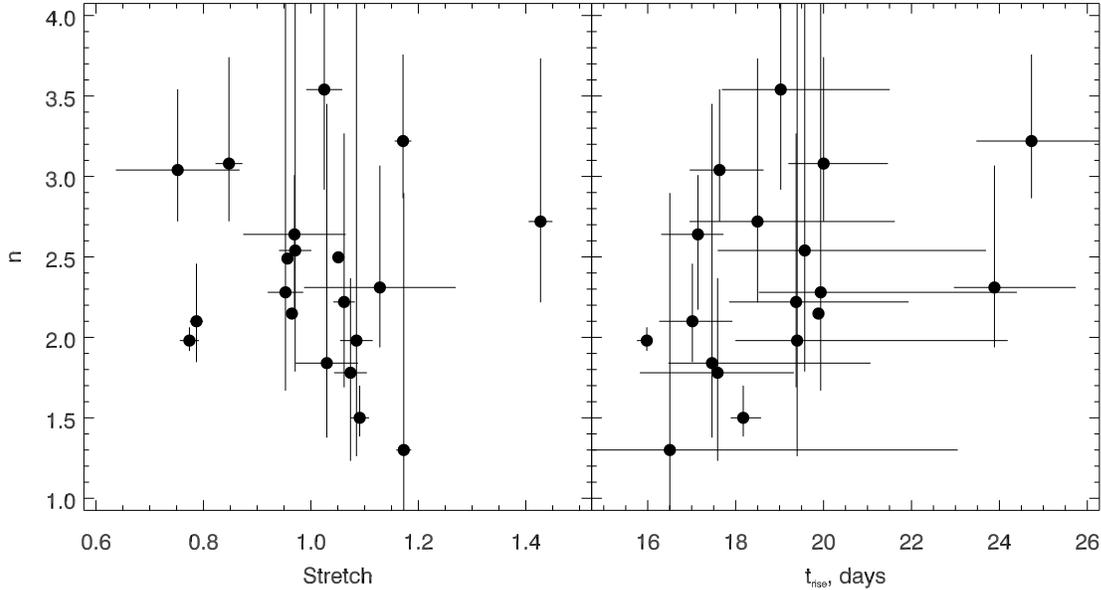}
	\caption{ \textit{Left Panel:} $n$ vs Stretch for our sample, no evidence of a correlation is found.
	\textit{Right Panel:} $n$ and \trise do show some evidence of a relationship, but with low significance.}
	\label{n_vs_tr_str}
\end{figure*}

We find no evidence of a correlation between $n$ and stretch (Fig \ref{n_vs_tr_str}, left panel).
\citet{RTSNLS2012} found a weak trend, with larger stretches 
corresponding with higher $n$.
While both \trise and Stretch do not correlate strongly with $n$, as can be seen in Fig \ref{n_vs_tr_str}, there is a clear correlation
between $t_{1}$ and $n$ (Figure \ref{n_v_t2_t1}, top panel), with the lowest
rise indices corresponding to the shortest initial time spans.

A distribution of $n$ values centred above $2$ agrees well with previous 
work on individually fitted SNe 
\citep{11FE2011,2013MNRAS.429.2228H,2013ApJ...778L..15Z,2014ApJ...783L..24Z}.
 However, these fits were done on rise time regions of varying sizes, it is for
 this reason that our value of $n$ for SN\,2011fe differs from that of \citet{11FE2011};
 in Fig. \ref{kly_cutoff}, when the data range fitted is the same, the values are fully 
 consistent. Thus direct comparisons between studies are difficult, as a shorter, earlier 
 fitting region probes a shallower ejecta region, raising the prospect of a time dependent
 index.

\subsubsection{A Time Dependent Index - $`\dot{n}$'}

Figure \ref{kly_cutoff} shows that the $n$ measured  changes over time,
as more data is added the behaviour of $n$ in Fig. (\ref{kly_cutoff}), is evidence that this is occurring. 
This effect explains the difference in $n$ measured in this work ($n=2.15\pm 0.02$,
 for $t_0 \leq t \leq t_{0.5}$) and \citet{11FE2011} ($2.01\pm0.01$ for $t_0 \leq t \leq t_0 + 3$)
procedure as outlined in Section \ref{sec:fitting-methods}, substituting eqn. \ref{ndot}
for eqn. \ref{eqn0}. 

We find evidence for a positive $\dot{n}$ in most SNe in our sample, with a mean value of 
$\dot{n} = 0.011\pm0.004\,d^{-1}$ (where the uncertainty is the standard error on the mean)
 and a weighted mean value of $\dot{n} = 0.011\pm0.001\,d^{-1}$. Specifically, in the case of SN\,2011fe with n 
as a free parameter we find $n_0 = 2.02\pm0.02$, consistent with \cite{11FE2011}, 
and an $\dot{n} = 0.011\pm0.001\,d^{-1}$. This positive $\dot{n}$ (in both cases) reflects that
 fitting equation \ref{eqn0} we find the mean $n$ greater than 2. We find some evidence
 that the SNe that have observations longest after explosion are those which, in general, have the 
 largest $\dot{n}$; that is, the largest rate of deviation away from the fireball model. This may be driven
  physically by the later time data being driven by deeper ejecta layers.
  
This time dependence of $n$ can also be seen in Fig \ref{n_vs_obsdate}. When the first observation
  of a SN is made earlier, the $n$ is lower, due to the different ejecta conditions. Applying a linear fit, an intercept of $n=1.8\pm0.2$ is found,
  and a slope of $m=0.39\pm0.15$, making the trend significant to $2.6\sigma$. When the observations
  begin at a later epoch, there is a smaller contribution from the \nickel in the upper most layers, changing
  the measured $n$.

\subsubsection{Broken Powerlaw}

We also performed a fit to SN\,2011fe using equation \ref{brokenpwr}. Unlike 
\citet{2013ApJ...778L..15Z} and \cite{2014ApJ...783L..24Z} we find no evidence 
for a break in the light curve. The data from SN\,2011fe contains 6 data points within
the first three days after the explosion (i.e. before a `break time'; $ t_{b}^{2013dy} = 3.14\pm0.30\,d$ and $t_{b}^{2014J} = 2.61\pm0.20\,d$),
however these are clustered in three epochs, and it may be that sub-day cadence is needed in this early time
to be sensitive to broken power laws.

\section{Discussion}
\label{sec:discussion}
In \citet{RPRL2012}, the first four days of data from SN\,2011fe were 
analysed and the implications of a power-law dependence explored,
considering the dynamics and thermodynamics of the expanding ejecta 
in shells. At early times, the emitted luminosity originates from a shell between
 $\simeq0.01$ and $0.3 M_{\odot}$. \citet{RPRL2012} calculate that the power-law scaling for the 
bolometric luminosity goes as $L(t) \propto t^{2(1+1/\gamma+\chi)/(1+1/\gamma+\beta)}$ for
 $t\la t_{56}$, with the polytropic index, $\gamma=3/2$ for non-relativistic
 electrons ($\gamma=3$ for relativistic electrons), and for a radiation 
 pressure dominated shock,  $\beta=0.19$ which controls the rate of change
  of the shock velocity,  while $\chi$ characterises the \nickel distribution in 
  the ejecta shell. To change the $n$ value, either $\chi$ or $\beta$ must 
  change. Simplifying the  expression by setting $\chi=0$ results in 
  $n\simeq1.8$. This value is consistent with \citet{RTSNLS2006}, 
  despite the fitting region being twice the size of the $4$ days for SN\,2011fe. 
  If we treat the region from $t_0$ to $t_{0.5}$ as one shell, as 
  in this parameterisation, increasing $n$ is possible by increasing $\chi$, 
  and having deeper \nickel dominating the rise. However smaller $n$ are more 
  problematic to explain.
  
It should be noted that, in Fig \ref{n_vs_obsdate}, the intercept of the best fit, 
  at $n=1.8$, is consistant with the above case from \citet{RPRL2012}. This value is in tension with the findings of higher $n$ values in SN\,2013dy 
  and SN\,2014J and the justification that the value of $n$ found was due to the unprecedented
  early discovery and followup. Clearly, data on further SNe collected very soon after first light are needed.
  
Only one of the SNe in our sample has $n<1.8$, PTF10accd, and the small uncertainties make it inconsistent with both $n=2$ and the 
  lower limit of \citet{RPRL2012} ($n=1.8$). It should be reiterated that the bolometric 
  value is expected to be larger than the values in \rptf\,or \gptf; however, from our 
  tests in Section \ref{sec:fits-bolom-vers} this would not make PTF10accd consistent
  with $n=1.8$.
   
 This result has two possible implications depending on $\chi$.
 If $\chi<0$, either \nickel dominates the makeup of the outer ejecta, or the flux originates from 
   elsewhere; potentially from some CSM interaction. Alternatively, 
   the optical luminosity of the shock-heated cooling light curve may be dominant 
   as this is expected to have $n=1.5$.
If $\chi\geq0$, then the shock is not radiation pressure dominated and $\beta$
may vary, or the delayed detonation transition (DDT) model, from which the velocity 
gradient is calculated \citep{2010ApJ...708..598P} is an incomplete description
of this process. Other models, such as He double-detonation 
\citep{2010A&A...514A..53F} or the collision model \citep{2013ApJ...778L..37K,2014arXiv1401.3347D}
present different treatments of the velocity gradient.
   
More recent work \citep{2013ApJ...769...67P} investigates the contribution 
of \nickel heating, both directly and from the diffusive tail, throughout the ejecta.
In the appendices of \citet{2013ApJ...769...67P}, rather than treating shells of material 
individually, integrals are evaluated over the entire ejecta. This leads to an altogether
more complex picture of the energy generation, which depends on the relative
fraction of \nickel throughout the ejecta, as given by

\begin{align}
	X_{56}(x) = \frac{1}{1+\exp[-\beta(x - x_{1/2})]},
\label{eq_piro}
\end{align}
\\
\begin{figure}
	\includegraphics[width=84mm]{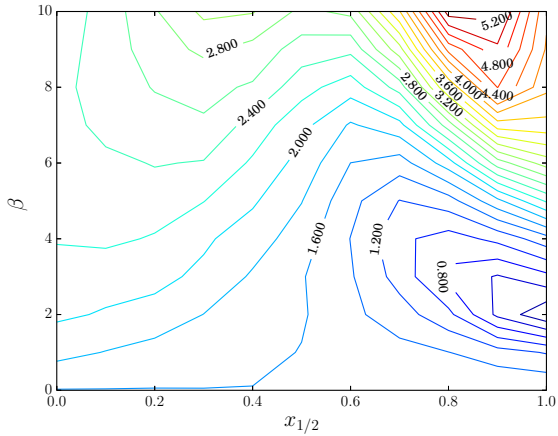}
	\caption{Contours showing the fitted $n$ parameter of a bolometric light curve
	generated by using different values of $\beta$ and $x_{1/2}$ in Equation \ref{eq_piro}.
	A large range of $n$ values are recovered, but extreme values of $\beta$ and $x_{1/2}$
	 may not be physical.}
	\label{b_xhalf}
\end{figure}

\noindent where $x$ is a measure of depth within the ejecta, $x_{1/2}$ is the point at which the 
\nickel fraction is half at that of its value at peak, and $\beta$ is the steepness of the rise 
in the distribution. Generating a bolometric luminosity using this parameterisation, 
it is possible to attain a large range of $n$ values, when fitting a power-law (Fig. \ref{b_xhalf}). 
The physical limits of this parameter space are uncertain, and fitting a number of SNe
 directly with this has yielded values in the ranges
 $6 \leq \beta \leq 8$ and $ x_{1/2} \simeq 0.9 $. The mean value of our study indicates that on 
 average the envelope is less well mixed and there is an abrupt change within the \nickel
 distribution in most cases in the sample.
 
 However, at very early times, fitting with a simple power law,
  the fit is poor, as the light curve is better described as an exponential 
  \citep{2014ApJ...784...85P}. The timescales for this discrepancy are short, and beyond
  the reach of this work. However, this may be the apparent `break' in the power law 
  seen in 2013dy and 2014J \citep{2013ApJ...778L..15Z,2014ApJ...783L..24Z} 
  - an exponential rise turning into a power law at later times. Our finding 
  that $\dot{n}$ is, in general, positive, supports this. 
  
  In Fig.~\ref{kly_cutoff} as the cutoff drops below $\sim3.5$ days, $n$
is consistent with 2, in agreement with \cite{11FE2011}. This value
differs from our final result for SN\,2011fe because the shorter time
period used only probes a shallow region of the ejecta. At very early
times, the rise index jumps to higher values. This difference
could be hinting towards a broken power-law as outlined in Section
2.1, Equation \ref{brokenpwr}, or something that resembles one
 \citep{2014ApJ...784...85P}; however, attempting to fit this model to
SN\,2011fe is unsuccessful; as there is not enough data at very early
times to constrain the 7 free parameters.

 At present, none of these various models make any predictions that would explain
the decoupling of the early and late part of the rise seen in our data. However, 
a scatter in the measured photometric rise time can be explained by invoking a
`dark phase', between explosion and first light (Fig \ref{labelled_template_color}),
 due to deep \nickel deposits. 
More work is needed to further understand this phenomena, using the methods
previously applied to SN\,2011fe and SN\,2010jn \citep{2014MNRAS.439.1959M,2013MNRAS.429.2228H}, 
on future samples of well observed SNe. The dark time for SN\,2010jn was estimated to be ~1.4 days,
 and that of SN\,2011fe to be ~1 day. As we expect that, for a given value of $x_{1/2}$, higher $n$ 
 values are consistent with a steeper gradient, $\beta$, higher $n$ values should be consistent
 with a longer dark time. This effect is seen in these two SNe; SN\,2011fe has an $n=2.15\pm0.02$ (our value is 
 used over that of \citet{11FE2011} as the fitted regions are more comparable) and SN\,2010jn
 $n=2.3\pm0.6$.

As the available light curves can be well fit by using a simple power law, 
more high quality photometric and spectroscopic data is needed to distinguish 
between the models, and to see expected deviations from power laws 
\citep{2013ApJ...778L..15Z,2014ApJ...784L..12G}. 
Ideally, future work would be able to concentrate on bolometric data,
which is now becoming possible \citep{2014MNRAS.440.1498S}.

\citet{2008OEIA} find a significant range of \nickel abundances in
the outer ejecta of a sample of SNe Ia, which is taken as one of the
causes of early time spectral variation in SNe Ia; they also suggest
this could have photometric consequences. Our work clearly
demonstrates that there is indeed a photometric shape variation, 
and that a cause of this is in \nickel deposition
between SNe Ia, as suspected. 

\subsection{Type Ia `CSM' Supernovae}

As considered in the previous subsection, one of the possible reasons for a 
SN to have an `anomalous' rise would be an energy contribution 
from interaction with CSM material \citep{1977ApJS...33..515F}. \citet{SilCSM2013} noted that the rise of 
SNe Ia-CSM tends to be significantly longer than a typical SNe Ia, following 
a simple photon diffusion argument - not only does a photon have to diffuse through
the ejecta, but also significant amounts of CSM as well. In that respect,
 \citet{2014ApJ...788..154O} showed that in SNe Type IIn there is a possible
 correlation between rise time and peak luminosity.
 
\begin{figure}
	\includegraphics[width=84mm]{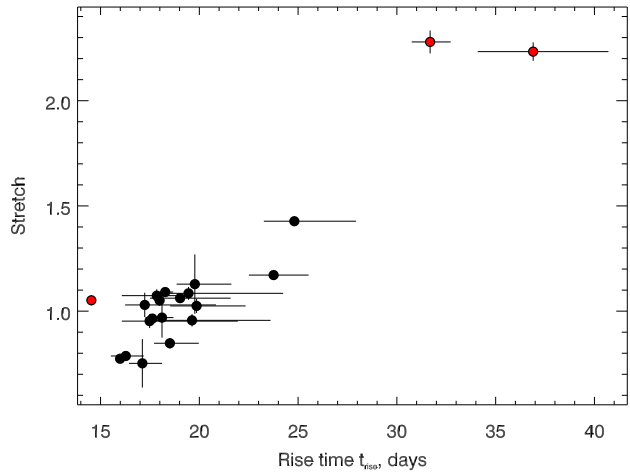}
	\caption{Stretch vs Rise time with 3 SNe Ia identified as CSM by \citet{SilCSM2013} plotted in red}
	\label{str_vs_tr_CSM}
\end{figure}

 We therefore examined  PTF SNe Ia-CSM 
from the sample of \citet{SilCSM2013} (7 SNe). Only 3 have 
sufficiently good photometry to provide acceptable fits, even after introducing 
constraints on the fitting. These fits assumed the fireball model,
 ($n=2$), and the results of measuring the rise time can be seen in Figure 
 \ref{str_vs_tr_CSM}. One of these SNe, PTF12efc is a `typical' broad and bright SNe-CSM, 
 although having an extreme rise time and stretch, and seems to lie in 
 agreement with the best fit to the distribution of normal SNe Ia. PTF10iuf also has a long rise and large stretch 
 but a higher stretch than would be predicted from the measured rise time. 
Despite lying on or near the correlation of `normal' SNe Ia, there seems to be no reason
for this to be the case - the rise is shaped by different physical processes over 
different timescales. 

\begin{figure}
	\includegraphics[width=84mm]{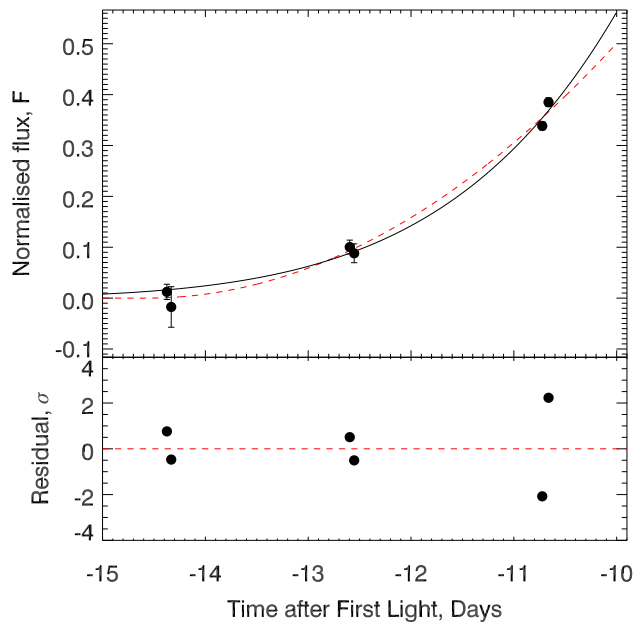}
	\caption{PTF11kx lightcurve and two fits to the data. A fit with fixed $n=2$ 
	is shown in red (dashed), and a fit with a fixed rise of $\trise=20.0$~days is shown in black.}
	\label{kx_LC}
\end{figure}

PTF11kx has a rise time of only 
$t_{r}^{n=2}=14.5\pm0.2$ (Fig. \ref{kx_LC}), but a measured stretch of $s=1.05$. In \citet{11KX2012}, 
a rise time of $\sim20$~days is assumed;  note that a shorter rise time means that the ejecta 
will be smaller at a given epoch. Consequently, the ejecta mass calculated using the previous estimate
($\sim5.3 \textnormal{M}_{\odot}$.) is too large, and should be $\sim 80\%$ of that value, 
$\sim4.3 \textnormal{M}_{\odot}$, making the same assumptions as in the supplemental information of
\citet{11KX2012}. As well as performing a fit holding $n=2$ a grid-search was done to find the best
 fit where $t_{r}=20$~days (Fig. \ref{kx_LC}). With a fixed 
 rise, the best fit index was $n=6.2\pm0.5$, this result is not physical for a `normal' 
 SN Ia,  but SNe Ia-CSM have an additional contribution to their light from the 
 collision of their ejecta with the CSM; this converts the kinetic energy in the ejecta 
 into hard X-ray photons, which in the presence of sufficient optical depth can be converted into
 optical light \citep[e.g.][]{2012ApJ...747L..17C,2012ApJ...759..108S,2014ApJ...788..154O}.
  With this in mind, it may be expected for SNe Ia-CSM to have abnormal 
 rise properties. 
 Until a confirmed SN Ia-CSM is observed with enough precision to enable a relaxing of $n$,
 few constraints can be placed on the effect of CSM on a rising light curve.
 
\section{Conclusions}

In this work we have used 18 type Ia supernovae (SNe Ia) from the
Palomar Transient Factory (PTF) and the La Silla-QUEST variability
surve (LSQ) to measure the rise time (\trise) (the time between first light $t_0$, and 
maximum light $t_{max}$) and rise index ($n$),
where $f= (t-t_0)^{n}$. Our main conclusions are:

\begin{enumerate}
  
\item The rise index, $n$, of our sample shows significant variation
  ($1.48\leq n \leq3.70$), with the mean of the distribution $n=2.44\pm0.13$,
  and $n=2.5\pm0.13$ when correcting to a pseudo-bolometric value 
  (Section~\ref{sec:rise-index-n}), both inconsistent with a simple fireball
  model ($n=2$) at a $3\sigma$ level.  This implies that current
  understanding of the \nickel distribution or shock velocity through
  the ejecta is incomplete, and that more complex physically motivated 
  parameterisations may be needed in future.

\item We find that when the rise index is allowed to vary with time from $n=2$,
$\dot n$ is in general positive, with an average value of $\dot n = 0.011\pm0.004\,$ day$^{-1}$. Supporting
 a time dependant $n$, is the finding that SN discovered later after first light have, in general
 a larger value of $n$, whereas those discovered soonest after $t_0$ have lower values.

\item The average \trise\ of our sample is
  $\trise=18.98\pm0.54$\,days, and $\trise=18.97\pm0.43$\,days, when
  correcting for light curve width. We find no evidence for two rise
  time modes in our sample. These are longer than would ordinarily be
  found by enforcing $n=2$.

\item The broadest light curves have a \trise\ that is faster than
  that of our stretch-corrected light curve template, which enforces
  $n=2$ in its construction. In agreement with previous studies, we
  find that a `two stretch' model fits the data better. In contrast to
  current two-stretch fitting methods, which separate the light curve
  into the pre- and post-maximum sections 
  ($t< t_{max}$ and $t>t_{max}$ respectively), the most significant
  variation occurs at the very earliest epochs ($t<t_{0.5}$, where
  $t_{0.5}$ is the time at which the SNe reaches half of its maximum, or
 phases $\tau<-10$\,d).
  
\item We therefore decouple the rise time into two components: $t_1$
  (where $t_1 = t_{0.5}-t_{0}$) and $t_2$ (where $t_2 = t_{max} - t_{0.5}$).
   These time-scales are not correlated
  with each other (Fig. \ref{n_v_t2_t1}); furthermore $t_2$ is strongly correlated
  with stretch, whereas $t_1$ is not. As a result, stretch correcting
  using a single stretch is ineffective in reducing the dispersion in
  the earliest portion of the light curve (Fig. \ref{STACK_LC}).

\item These two regions are separated by the approximate location of
  the point at which energy deposition and radiation are equal,
  meaning that the physical conditions are distinct.
   
\item Using models from \cite{2013ApJ...769...67P}, we show that 
potential variation in the shape of the \nickel\ distribution within the 
SN ejecta can explain the measured range of $n$ and \trise.

\item SNe Ia showing evidence of strong interaction with circumstellar
material (CSM) have long rise times. However a notable member of this
subclass, PTF11kx, has an extremely short rise, $\trise = 14.5\pm0.5$ days,
when fitted with a fireball model ($n=2$).
  
\end{enumerate}

Further work should concentrate on further understanding the variation, and on which
other observable quantities it depends. For this, a large sample of SNe Ia with high
quality photometric and spectroscopic data must be assembled. The presence of 
high velocity features, Si II velocities and colour evolution may hold valuable information,
particularly if the variation at very early times is misunderstood when used for cosmology.

\section{Acknowledgements}

We wish to thank the Reviewer for a careful and considered reading, and thorough
and helpful report.

MS acknowledges support from the Royal Society and EU/FP7-ERC grant no [615929].

KM is supported by a Marie Curie Intra European Fellowship, 
within the 7th European Community Framework Programme (FP7). 

This research used resources of the National Energy Research Scientific Computing 
Center, which is supported by the Office of Science of the U.S. Department of Energy 
under Contract No. DE-AC02-05CH11231.

E.O.O. is incumbent of the Arye Dissentshik career development chair and
is grateful to support by grants from the Willner Family Leadership Institute
Ilan Gluzman (Secaucus NJ), Israeli Ministry of Science, Israel Science Foundation,
Minerva, Weizmann-UK and the I-CORE Program of the Planning
and Budgeting Committee and The Israel Science Foundation.

Observations obtained with the Samuel Oschin Telescope at the Palomar
Observatory as part of the Palomar Transient Factory project, a
scientific collaboration between the California Institute of
Technology, Columbia University, Las Cumbres Observatory, the Lawrence
Berkeley National Laboratory, the National Energy Research Scientific
Computing Center, the University of Oxford, and the Weizmann Institute
of Science. 

Some of the data presented herein were obtained at the
W.M. Keck Observatory, which is operated as a scientific partnership
among the California Institute of Technology, the University of
California and the National Aeronautics and Space Administration. The
Observatory was made possible by the generous financial support of the
W.M. Keck Foundation.

This research was made possible through the use of the AAVSO Photometric 
All-Sky Survey (APASS), funded by the Robert Martin Ayers Sciences Fund.

\bibliography{WorkRead}

\begin{center}
\begin{table*}{}
\centering
 \caption{Table of example data for one SN, PTF09dsy. Data for the full sample can be found in online supplemental material}
 \label{data_tab}
 \begin{tabular}{@{}lcccccc}
 \hline
  SN & MJD & Counts & $\Delta$ Counts & Filter & Zero Point & Redshift\\
  \hline
  PTF09dsy & & & & & &\\
  \hline
 & 55054.441  &   -67.7  &   98.2 & \rptf & 27.000 & $0.0131\pm0.001$ \\   
 & 55054.460  &   -68.7  &   72.3 & \rptf & 27.000 & $0.0131\pm0.001$ \\   
 & 55055.460  &   480.0  &  101.0 & \rptf & 27.000 & $0.0131\pm0.001$ \\   
 & 55055.472  &   457.2  &   67.6 & \rptf & 27.000 & $0.0131\pm0.001$ \\   
 & 55059.443  &  8599.9  &   93.5 & \rptf & 27.000 & $0.0131\pm0.001$ \\   
 & 55059.463  &  8603.4  &   94.4 & \rptf & 27.000 & $0.0131\pm0.001$ \\   
 & 55061.434  & 16360.3  &  142.9 & \rptf & 27.000 & $0.0131\pm0.001$ \\   
 & 55061.459  & 16634.9  &  104.1 & \rptf & 27.000 & $0.0131\pm0.001$ \\   
 & 55063.438  & 27290.6  &  123.8 & \rptf & 27.000 & $0.0131\pm0.001$ \\   
 & 55063.483  & 26937.5  &  143.1 & \rptf & 27.000 & $0.0131\pm0.001$ \\   
 & 55066.421  & 38663.8  &  301.7 & \rptf & 27.000 & $0.0131\pm0.001$ \\   
 & 55066.466  & 39415.6  &  192.0 & \rptf & 27.000 & $0.0131\pm0.001$ \\   
 & 55069.410  & 46253.4  &  201.1 & \rptf & 27.000 & $0.0131\pm0.001$ \\   
 & 55069.455  & 46201.3  &  209.2 & \rptf & 27.000 & $0.0131\pm0.001$ \\   
 & 55080.373  & 30495.7  &  222.8 & \rptf & 27.000 & $0.0131\pm0.001$ \\   
 & 55080.461  & 29730.6  &  248.6 & \rptf & 27.000 & $0.0131\pm0.001$ \\   
 & 55087.350  & 24598.7  &  169.6 & \rptf & 27.000 & $0.0131\pm0.001$ \\   
 & 55087.395  & 24978.9  &  178.0 & \rptf & 27.000 & $0.0131\pm0.001$ \\   
 & 55089.381  & 23637.7  &  162.0 & \rptf & 27.000 & $0.0131\pm0.001$ \\   
 & 55089.425  & 23756.2  &  197.7 & \rptf & 27.000 & $0.0131\pm0.001$ \\   
 & 55093.351  & 20758.2  &  147.4 & \rptf & 27.000 & $0.0131\pm0.001$ \\   
 & 55093.395  & 20267.0  &  159.0 & \rptf & 27.000 & $0.0131\pm0.001$ \\   
 & 55095.338  & 18196.5  &  117.7 & \rptf & 27.000 & $0.0131\pm0.001$ \\   
 & 55095.382  & 18400.1  &  168.8 & \rptf & 27.000 & $0.0131\pm0.001$ \\   
 & 55107.311  &  8207.5  &  146.7 & \rptf & 27.000 & $0.0131\pm0.001$ \\   
 & 55107.355  &  8001.8  &  136.7 & \rptf & 27.000 & $0.0131\pm0.001$ \\  
 \hline
 \end{tabular}
\end{table*}
\end{center}

\begin{landscape}
\begin{center}
\begin{table}
\centering
 \caption{Table of Results}
 \label{results_tab}
 \begin{tabular}{@{}lcccccccccc}
  \hline
  SN & $t_{max}, $ & \trise,   & $n$ & $\dot{n}$ & Stretch & $\chi^{2}_{\textnormal{DOF}}$ & R.A & Dec. & Filter \\
  & \textnormal{MJD}& days& & & & & (J2000) & (J2000) & \\
  \hline
PTF09dsy &        $55070.4\pm0.1$ & $      15.98\pm0.20 $ & $      2.00^{+    0.08} _{-    0.07} $ & $0.0^{+0.13}_{-0.01}$ & $     0.80\pm    0.01$ &      0.58 & 3:33:22.1 & -04:59:55.2 & PTF48R  \\
PTF10accd &        $55556.0\pm0.2$ & $      18.17^{+     0.46} _{-     0.32} $ & $  1.48^{+     0.19} _{-     0.12} $& $-0.02\pm0.01$ & $      1.11\pm    0.02$ &      0.78	& 02:13:30.4 & 46:41:37.2 & PTF48R \\
PTF10duz &        $55285.0\pm0.2$ & $      17.5^{+      3.7} _{-      1.0} $ & $      1.96^{+      1.5} _{-     0.46} $ & $-0.005^{+0.02}_{+0.01}$&$      1.00\pm    0.03$ &       1.88 & 12:51:39.5 & 14:26:18.7 & PTF48R \\
PTF10hml &        $55352.3\pm0.1$ & $      19.4^{+      2.7} _{-      1.5} $ & $      2.31^{+      1.08} _{-     0.53} $ &$0.01^{+0.01}_{-0.02}$ &$      1.07\pm    0.02$ &       1.50 & 13:19:49.7 & 41:59:1.6 & PTF48R\\
PTF10iyc &        $55361.5\pm0.1$ & $      19.4^{+      4.4} _{-      1.4} $ & $      2.08^{+      2.00} _{-     0.70} $ & $0.00^{+0.02}_{-0.03}$&$      1.10\pm    0.02$ &       1.04 & 17:09:21.8 & 44:23:35.9 & PTF48R \\
PTF11gdh &        $55744.1\pm0.1$ & $      19.57\pm1.8$ & $      2.26^{+     0.58} _{-     0.55} $ & $-0.005\pm0.01$&$      1.07\pm    0.03$ &       1.25 & 13:00:38.1 & 28:03:24.1 & PTF48R \\
PTF11hub &        $55770.0\pm0.2$ & $      16.50^{+     0.96} _{-     0.76} $ & $      2.17^{+     0.35} _{-     0.26} $ & $0.005\pm0.01$& $     0.80\pm   0.01$ &       1.17 & 13:12:59.5 & 47:27:40.3 & PTF48R \\
PTF11kly/SN2011fe&        $55814.3\pm0.1$ & $      17.59\pm0.1 $ & $      2.15\pm0.02 $ &$0.011\pm0.001$ & $     0.965\pm   0.009$ & $8.06 $ & 14:30:5.8 & 54:16:25.2 & PTF48g \\
PTF11qnr &        $55902.3\pm0.1$ & $      17.01^{+      1.0} _{-     0.7} $ & $      2.99^{+     0.49} _{-     0.32} $ & $0.025^{+0.01}_{-0.005}$& $     0.79\pm    0.04$ &       1.57 & 22:44:25.4 & -00:10:2.0 & PTF48R \\
PTF12emp &       $ 56080.9\pm0.4$ & $      19.9^{+      1.9} _{-      1.0} $ & $      2.36^{+     0.76} _{-     0.37} $ & $0.01^{+0.01}_{-0.01}$& $      1.13\pm     0.14$ &       3.52 & 13:13:53.7 & 34:06:59.7 & PTF48R \\
LSQ12fxd &        $56246.4\pm0.1$ & $      23.8^{+      1.8} _{-      1.3} $ & $      3.24^{+     0.53} _{-     0.36} $ & $0.02^{+0.02}_{-0.2}$& $      1.17\pm    0.01$ &       2.77 & 05:22:17.0 & -25:35:47.0 & LSQgr \\
PTF12gdq &        $56116.3\pm1.8$ & $      17.6^{+      4.7} _{-      2.3} $ & $      2.34^{+      1.86} _{-     0.61} $ & $0.015\pm0.02$& $     0.94\pm    0.02$ &       2.06 & 15:11:35.3 & 09:42:34.0 & PTF48R \\
LSQ12gpw &        $56268.4\pm0.1$ & $      24.7^{+      3.2} _{-      1.6} $ & $      2.74^{+     1.00} _{-     0.50} $ &$0.015\pm0.01$ & $      1.42\pm    0.02$ &       5.79 & 03:12:58.2 & -11:42:40.0 & LSQgr \\
LSQ12hxx &        $56289.8\pm0.1$ & $      19.9^{+      2.3} _{-      1.3} $ & $      3.70^{+      1.08} _{-     0.61} $ & $0.04\pm0.01$& $      1.00\pm    0.03$ &       2.00 & 03:19:44.2 & -27:00:25.6 & LSQgr \\
LSQ12hzj &        $56300.8\pm0.2$ & $      18.5^{+     0.6} _{-     0.8} $ & $      2.61^{+     0.37} _{-     0.47} $ & $0.045^{+0.02}_{-0.03}$& $ 0.97\pm    0.05$ &       20.19 & 09:59:12.4 & -09:0:8.30 & LSQgr \\
LSQ13ry &        $56394.9\pm0.1$ & $      19.0^{+      1.5} _{-     0.8} $ & $      3.03^{+     0.64} _{-     0.37} $ & $0.025\pm0.01$& $     0.86\pm    0.02$ &       1.81 & 10:32:48.0 & 04:11:51.4 & LSQgr \\
LSQ13cpk &        $556590.0\pm0.1$ & $      17.01^{+      0.15} _{-     0.25} $ & $      1.87^{+     0.07} _{-     0.13} $ &$0.02^{+0.04}_{-0.2}$ & $     1.05\pm    0.03$ &       1.77 & 02:31:3.8 & -20:08:49.6 & LSQgr \\
LSQ13des &        $56638.9\pm0.1$ & $      20.0^{+      3.9} _{-     2.3} $ & $      2.64^{+     1.37} _{-     0.8} $ & $0.01\pm0.01$ & $     0.96\pm    0.03$ &       2.79 & 03:25:18.9 & -23:42:3.5 & LSQgr \\
 \\  
 \hline
 \end{tabular}
\end{table}
\end{center}
\end{landscape}

\appendix

\label{lastpage}

\end{document}